\documentclass[a4paper,11pt]{article}
\pdfoutput=1

\usepackage{jheppub}

\usepackage[final]{pdfpages}
\usepackage[numbers,sort&compress]{natbib}

\usepackage{graphicx}
\RequirePackage{doi}
\usepackage{multirow}
\usepackage{amsmath}
\usepackage{slashed}
\usepackage{float}
\usepackage{rotating}

\usepackage{xcolor}
\usepackage{subcaption}
\usepackage{orcidlink}
\graphicspath{ {./images/} }

\newcommand{\Amp}{\mathcal{M}}
\newcommand{\Ham}{\mathcal{H}}

\newcommand{\mycomment}[1]{}
\newcommand{\cng}[1]{\textcolor{black}{#1}}

\def\BDlnu{$ {B} \to D^*\ell {\nu}_\ell \;$}
\newcommand{\BDslnu}{ \bar{B} \to D^*\ell {\nu}_\ell }
\newcommand{\BDslnudpi}{ \bar{B} \to D^* (\to D \pi)\ell {\nu}_\ell }
\newcommand{\BDstaunu}{ \bar{B} \to D^* \tau {\bar{\nu}}_\tau }

\newcommand{\BDstaufull}{ \bar{B} \to D^* (\to D \pi) \tau (\to \ell \nu_\tau \bar{\nu}_\ell) \bar{\nu}_\tau}
\newcommand{\tlnn}{\tau \to \ell \nu_\tau \bar{\nu}_\ell }
\newcommand{\linemu}{\ell \in \{ e , \mu \}}
\newcommand{\Dsdpi}{ D^* \to D \pi }
\newcommand{\eds}{\epsilon_{D^*}}
\newcommand{\lds}{{\lambda_{D^*}}}
\newcommand{\ew}{\epsilon_{W}}
\newcommand{\ra}{\rightarrow}

\def\BDtaunu{\bar{B} \to D \tau^{-} {\bar\nu}_\tau}
\def\BDlnu{\bar{B} \to D \ell^{-} {\bar\nu}_\ell}
\def\BDstartaunu{\bar{B} \to D^{*} \tau^{-} {\bar\nu}_\tau}
\def\BDstarlnu{\bar{B} \to D^{*} \ell^{-} {\bar\nu}_\ell}

\def\bsmumu{b \to s \mu^+ \mu^-}

\def\bsee{b \to s e^+ e^-}
\def\bsll{b \to s \ell^+ \ell^-}

\def\RDns{R({D^{*}})}
\def\RDn{R({D})}

\def \bea{\begin{eqnarray*}}
\def \beq{\begin{equation}}

\def \eea{\end{eqnarray*}}
\def \eeq{\end{equation}}

\def \({\left(}
\def \){\right)}
\def \[{\left[}
\def \]{\right]}
\def \bma{\begin{matrix}}
\def \ema{\end{matrix}}

\newcommand{\cn}[1]{\textcolor{black}{#1}}

\usepackage{hyperref}

\title{New physics searches via angular distributions of $\BDstaufull$ decays}

\author[a]{Bhubanjyoti Bhattacharya\,\orcidlink{0000-0003-2238-321X},}
\author[b]{Thomas E. Browder,}
\author[c]{Alakabha Datta,}
\author[d]{Tejhas Kapoor\,\orcidlink{0000-0001-5726-3037},}
\author[d]{Emi Kou,}
\author[e]{and Lopamudra Mukherjee\,\orcidlink{0000-0001-8765-7563}}

\affiliation[a]{Department of Natural Sciences, Lawrence Technological University, Southfield, MI 48075, USA}
\affiliation[b]{Department of Physics and Astronomy, University of Hawaii, Honolulu, HI 96822, USA}
\affiliation[c]{Department of Physics and Astronomy, 108 Lewis Hall, University of Mississippi, Oxford, MS 38677-1848, USA}
\affiliation[d]{Universit\'e Paris-Saclay, CNRS/IN2P3, IJCLab, 91405 Orsay, France}
\affiliation[e]{School of Physics, Nankai University, Tianjin 300071, China}

\emailAdd{bbhattach@ltu.edu}
\emailAdd{teb@phys.hawaii.edu}
\emailAdd{datta@phy.olemiss.edu}
\emailAdd{tejhas.kapoor@etu-upsaclay.fr}
\emailAdd{emi.kou@ijclab.in2p3.fr}
\emailAdd{lopamudra.physics@gmail.com}

\abstract{
The study of $\BDstaunu$ angular distribution can be used to obtain information about new physics (or beyond the Standard Model) couplings, which are motivated by various $B$ anomalies. However, the inability to measure precisely the three-momentum of the $\tau$ lepton hinders such measurements, as the tau decay contains one or more undetected neutrinos. Here, we present a measurable angular distribution of $\BDstaunu$ by considering the additional decay $\tlnn$, where $\linemu$. The full process used is $\BDstaufull$, in which only the $\ell$ and $D^*$ are reconstructed. A fit to the experimental angular distribution of this process can be used to extract information on new physics parameters. To demonstrate the feasibility of this approach, we generate simulated data for this process and perform a sensitivity study to obtain the expected statistical errors on new physics parameters from experiments in the near future. We obtain a sensitivity of the order of 5\% for the right-handed current and around 6\% for the tensor current. In addition, we use the recent lattice QCD data on $B \to D^*$ form factors and obtain correlations between form factors and new physics parameters.}

\begin{document}

\maketitle
\flushbottom

\section{Introduction}
Among its many quirks, the flavor structure of the Standard Model (SM) remains the least understood. There is no understanding of the quark and lepton masses and mixings, including CP violation. This suggests that new physics (NP) must be present to explain the flavor structure of the SM. Additional states like sterile neutrinos or dark sector particles can enlarge the flavor sector. Results from various flavor physics experiments can give clues to the underlying flavor structure of the SM and of NP through observations of virtual effects from new states in various processes.

\begin{figure}[htb!]
   \includegraphics[width=14.0 cm]{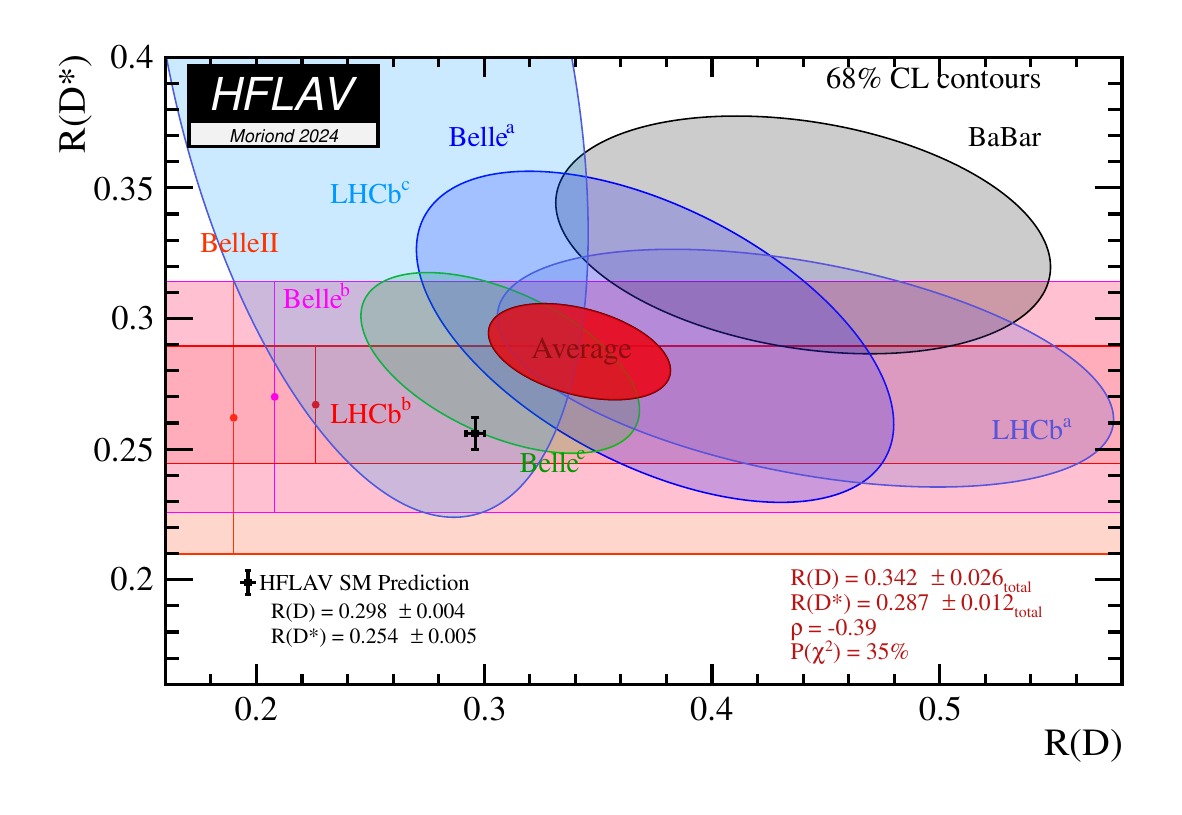}~~\\
  \caption{\small  68\% confidence limit (CL) contours of $R(D)-R(D^*)$ values, as measured by different experiments(BaBar~\cite{lfubabar1}~\cite{lfubabar2}, Belle~\cite{lfubellea}\textsuperscript{a}~\cite{lfubelleb1}\textsuperscript{b}~\cite{lfubelleb2}\textsuperscript{b}~\cite{lfubellec}\textsuperscript{c}, Belle II~\cite{lfubelleii} and LHCb~\cite{lfulhcba}\textsuperscript{a}~\cite{lfulhcbb}\textsuperscript{b}~\cite{lfulhcbc}\textsuperscript{c}). The experimental values of $R(D)$ and $R(D^*)$ exceed the SM prediction by about $1.7\sigma$ and $2.5\sigma$, respectively.}
   \label{RDpuzzle}
 \end{figure}

One of the outstanding puzzles of the flavor structure of the SM is the existence of three families of quarks and leptons. The quarks and leptons are produced as flavor eigenstates and one has to make unitary transformations to the mass basis. In the mass basis, tree-level flavor changing charged current transitions occur but there are no tree-level Flavor Changing Neutral Current (FCNC) processes. A crucial assumption behind this result is that the gauge bosons in the SM couple equally to the three generations of quarks and leptons. This assumption has to be tested. The assumption of universal gauge interactions of the leptons is in tension with results in semileptonic $B$ decays known as the $B$ anomalies. The $B$ anomalies appear in both the charged and neutral current semileptonic $B$ decays. In the charged current decays, 
the processes $\BDlnu$ and $\BDstarlnu$ with the leptons $\ell= e, \mu, \tau$ are simple tree-level processes in the SM. There can be tree-level NP contributions in these decays through the exchange of new mediators such as the charged Higgs \cite{h1,h2,h3,h4,h5,h6,h7,h8,h9,h10,Datta:2023mmb}, extra gauge bosons \cite{w1,w2,w3,w4,w5,w6} and leptoquarks \cite{tensorff,lq1,lq2,lq3,lq4,watanbeglobalfit}.
Lepton flavor universality of the gauge interactions can be tested through the measurements of the following ratios
\begin{eqnarray*}
\label{babarnew}
R(D) &\equiv& \frac{{\cal B}(\BDtaunu)}
{{\cal B}(\BDlnu)} \quad \quad
R(D^*) \equiv \frac{{\cal B}(\BDstartaunu)}
{{\cal B}(\BDstarlnu)}.
\label{RDexpt}
\end{eqnarray*}
Over the years, measurements of these ratios \cite{babaranomaly1, babaranomaly2, belleanomaly1, belleanomaly2, belleanomaly3, belleanomaly4, belle2anomaly1, lhcbanomaly1,lhcbanomaly2} have shown deviations from SM expectations \cite{hflav} as shown in Figure.\ref{RDpuzzle}.
Including correlations, one finds that the deviation is at the level of $3.3 \sigma$ from the SM expectation. This is known as the $\RDn- \RDns$ puzzle and is one of the main motivations for this work.

However, to provide a better context for the anomalies, it is also pertinent to discuss the neutral current anomalies. Over the last decade, there have been several measurements of branching ratios and angular observables of the semileptonic decay $\bsll$ ($\ell = \mu, e$) \cite{rkanomaly1,rkanomaly2,rkanomaly3,rkanomaly4,rkanomaly5,rkanomaly6,rkanomaly7,rkanomaly8,rkanomaly9,rkanomaly10}, which are in disagreement with SM predictions~\cite{hflav}.  Initially, it appeared that only the $\bsmumu$ decays were affected by NP and this was clear evidence of lepton universality violating NP. However, updated measurements by LHCb of the ratios $R(K)$ and $R(K^*)$, which test for lepton-flavor universality, now agree with the SM~\cite{LHCb:2022qnv, LHCb:2022vje}. At this point, though the branching ratios and angular observables for $\bsmumu$ processes are still discrepant from the SM, the lepton universality violating (LUV) ratios seem to be consistent with the SM, indicating that NP, if present, is lepton universal at least in the muon and electron sectors. Note that the branching ratios and angular observables for $\bsmumu$ processes can be explained by charm loop contributions, though this framework also has its theoretical challenges and can be tested with additional data.
Now, a promising NP explanation of the neutral current anomalies is that the NP contributes equally to $\bsmumu$ and $\bsee$ but has its source in nonuniversal NP, including LUV NP connected to the third generation with RGE effects producing equal contributions to $\bsmumu$ and $\bsee$~\cite{Greljo:2022jac, Alguero:2023jeh, Hurth:2023jwr, Datta:2024zrl}. A model to generate equal contributions to $\bsmumu$ and $\bsee$ from NP affecting third-generation physics operators was proposed much earlier~\cite{Datta:2013kja}. In many models, NP in the charged current and neutral current semileptonic decays are connected (see~\cite{Bhattacharya:2014wla, Bhattacharya:2016mcc}) and so a clear understanding of the
$\RDn- \RDns$ puzzles will likely also have implications for the anomalies in the neutral current sector.

$\BDtaunu$ and $\BDstartaunu$ are tree-level decays in the SM. Therefore, they have larger branching ratios than loop-induced FCNC decays. Furthermore, it has been shown that observables in angular distributions, such as polarization fractions or angular asymmetries, are useful tools for testing theoretical predictions of form factors \cite{Fedele:2023ewe} and distinguishing between the SM and various types of NP scenarios \cite{Blanke:2018yud,Blanke:2019qrx,Bhattacharya:2022bdk}. Increased statistics from experiments will allow one to measure the angular distributions in these decays with more data and, consequently, more sensitivity to NP effects. One of the issues with a $\tau$ lepton in the final state is that, unlike the light leptons, the $\tau$ is reconstructed from its decay products. Since the decay of the $\tau$ involves a neutrino, there are at least two neutrinos in the final state from the $B$ decay. Thus, the $\tau$ rest frame cannot be fully reconstructed. Hence, the angular distributions have to be expressed in terms of the visible daughter states from the $\tau$ decay. \cn{A step in this direction was taken in Refs.~\cite{Nierste:2008qe,bdltaunudatta} where the two-body $\tau$ decay $ \tau^- \to \pi^- \nu_\tau$ was considered. Similarly, \cite{ Alonso:2016gym} started the work for the $\tlnn $ case. In this article, building upon the previously mentioned works}, we extend our analysis to the three-body leptonic decay of the $\tau$, which is $\tlnn$, where $\linemu$. The $\ell$'s four-momentum is measured in the $W$ rest frame, as the $\tau$ rest frame cannot be reconstructed.  In the following, unless explicitly stated, $\ell$ denotes both $e$ and $\mu$. 

The procedure we follow starts from an effective theory description of NP in terms of dimension-six operators, using which we work out the angular distribution in terms of the kinematic variables of the visible final states. We assume that the $\tau$ decay is not affected by NP. We derive the analytic expression for the angular distribution. We also present sensitivity studies for NP Wilson coefficients \cn{by performing fits on data simulated by a toy Monte-Carlo method, described in Appendix~\ref{app:statprocedure}}. Global fits have been performed to obtain constraints on right-handed, pseudoscalar, and tensor currents, where observables such as $R(D^{(*)})$, $D^{(*)}$ polarization fraction, $\mathcal{B}(B_c \to \tau\bar{\nu})$ etc are used: for example, see Refs.~\cite{globalfit1,globalfit2}. Our analysis provides sensitivities of the same order as in these global fits and will improve as experimental data becomes available. 
\par
The paper is organized in the following manner. In Section~\ref{sec:effectivehamiltonian}, we describe the NP effective Hamiltonian. In Section \ref{sec:angdistfullbdtaunu}, we calculate the amplitude and phase space to derive an analytical expression for the angular distribution of $\BDstaufull$. In Section \ref{sec:unbinnedanalysis}, we describe the generation of simulated data and perform a sensitivity study to obtain the expected sensitivities to NP couplings based on likelihood analyses from future experiments. Finally, in Section \ref{sec:conclusions} we summarize our work and present our conclusions.

\section{Effective Hamiltonian}\label{sec:effectivehamiltonian}

We start with the most general low energy effective Hamiltonian at the $m_b$ scale for $b \to c \ell \nu_\ell$ transition containing all dimension six, four-fermion operators, assuming no right-handed neutrinos{\footnote {Sometimes a slightly different form of the effective Hamiltonian is used -- see Ref.~\cite{Datta:2012qk}}}: 
\begin{align}\label{eq:effective_hamiltonian}
\Ham_{eff}= \frac{4G_F}{\sqrt{2}} V_{cb} [ C^{\ell}_{V_L} O^{\ell}_{V_L} + C^{\ell}_{V_R} O^{\ell}_{V_R} + {C^{\ell}_{S} O^{\ell}_{S}}+C^{\ell}_{P} O^{\ell}_{P} + C^{\ell}_{T} O^{\ell}_{T}],
\end{align}
where ${V_L, V_R, S, P, T}$ indicate the left-handed, right-handed, scalar, pseudoscalar, and tensor terms, respectively.  The corresponding operators are given as
\begin{align} 
\begin{aligned}
O^{\ell}_{\rm V_L} &= (\bar{c}_L \gamma^{\mu} b_L)(\bar{{\ell}}_L \gamma_{\mu} \nu_{{\ell}L}) \\
O^{\ell}_{\rm V_R} &= (\bar{c}_R \gamma^{\mu} b_R)(\bar{{\ell}}_L \gamma_{\mu} \nu_{{\ell}L}) \\
O^{\ell}_{\rm S} &=  (\bar{c}  b)(\bar{{\ell}}_R  \nu_{{\ell}L}) \\
O^{\ell}_{\rm P} &= (\bar{c} \gamma^5 b)(\bar{{\ell}}_R  \nu_{{\ell}L}) \\
O^{\ell}_{\rm T} &= (\bar{c}_R \sigma^{\mu\nu} b_L)(\bar{{\ell}}_R \sigma_{\mu\nu}  \nu_{{\ell}L}) 
\end{aligned}
\end{align}
where $\psi_{L/R}=(1\mp\gamma_5)\psi/2$ with $\psi=\{c, b, \ell, \nu\}$. In the SM, $C_{V_L} = 1$, while the remaining Wilson coefficients are zero. We ignore the superscript $\ell$ in the following discussion.


\section{Calculation of the angular distribution of $\BDstaufull$ decay}\label{sec:angdistfullbdtaunu}

In this section, we detail the calculation of the angular distribution of the $\BDstaufull$ decay by following a similar calculation done for the case of $\bar{B} \to D^* (\to D\pi) \tau (\to \pi \nu_\tau) \bar{\nu}_\tau$ \cite{bdltaunudatta}. As discussed below, the challenge of studying the angular distribution of this decay comes from the fact that the $\tau$ decay contains one or more neutrinos. Therefore, we cannot precisely measure the three-momentum (magnitude and direction) of the $\tau$. Consequently, we cannot measure the helicity angles of its decay products in its rest frame, as the rest frame of the $\tau$ is unknown. Here, we attempt to solve this issue by moving from the $\tau$ rest frame to the $W$ rest frame in order to measure the $\ell$ angles, where this $\linemu$ is the only detectable particle coming from the $\tau$ decay. The $B$-meson decay rate is given by
\begin{align}\label{eq:decayrategeneral}
d\Gamma = \frac{1}{2 M_B} \int d\Pi_n |\Amp|^2,
\end{align}
where $\Amp$ is the decay amplitude, presented in Section~\ref{sec:amplitudecalculation}, and $d\Pi_n$ is the $n$-body phase space element, detailed in Section~\ref{sec:bdtaunuphasespace}. Using all these elements, we give the full measurable angular distribution for the $B$ decay by integrating over non-observable quantities in Section~\ref{sec:angdistexp}.


\subsection{Amplitude calculation}\label{sec:amplitudecalculation}

\begin{figure}[t]
\setlength{\unitlength}{1mm}
  \centering
  \begin{picture}(140,60)
    \put(0,-1){
      \includegraphics*[width=130mm]{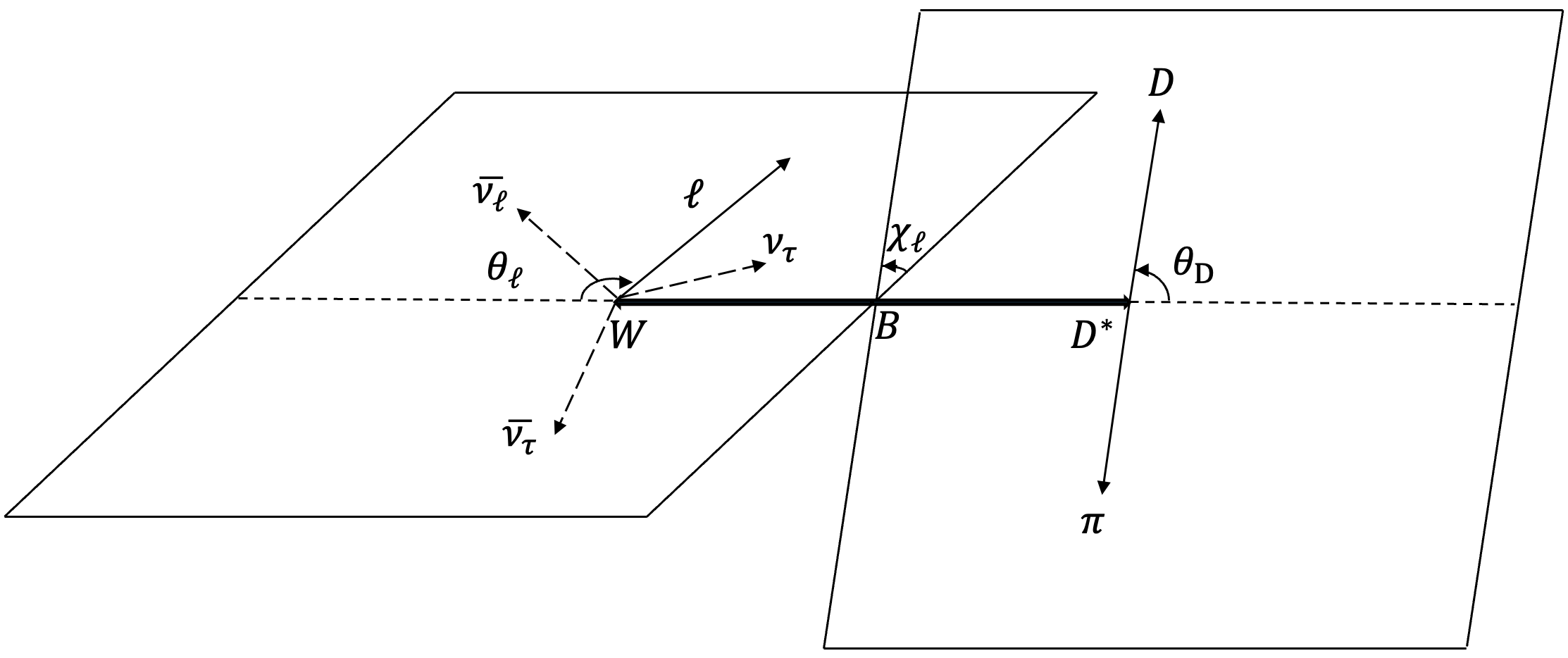}
    }
  \end{picture}
  \caption{\small  Decay angles for the $\BDstaunu$ decay, where $\theta_D$ is the angle between the direction of the $D$ meson and the direction opposite to that of $B$ meson, in the $D^*$ meson rest frame.  Similarly, $\theta_\ell$ is the angle between the direction of $\ell$ and the direction opposite to that of $B$ meson in the virtual $W$ rest frame.  $\chi_\ell$ is the angle between two decay planes. The plane on the left is formed by $\ell$ and $B$ vectors in the $W$ rest frame, while the one on the right is formed by the $D$ and $B$ vectors in the $D^*$ rest frame.  Note that, the $W$ first decays to a $\tau$ and a $\bar{\nu}_\tau$. The $\tau$ subsequently decays to $\ell \bar{\nu}_{\ell}\nu_{\tau}$. The angles of the $\tau$ three-momentum are not observable and hence are not shown. Yet, they appear in the intermediate computation steps and need to be integrated away to obtain an observable angular distribution.}
\label{fig:decayanglestau}
\end{figure}

\begin{figure}[t]
\setlength{\unitlength}{1mm}
\centering
\includegraphics*[width=80mm]{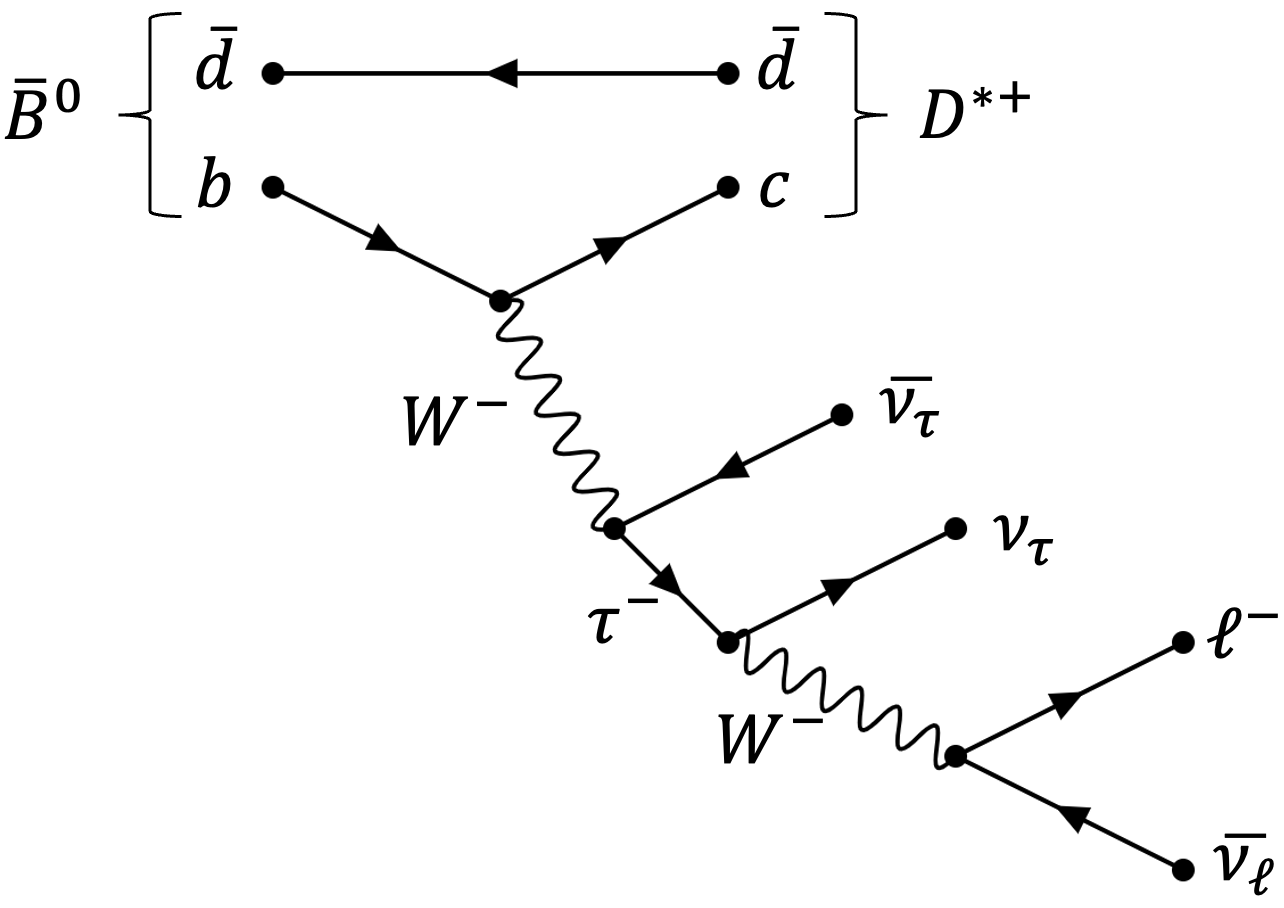}
\caption{\small Feynman diagram for $b \to c \tau \bar{\nu}_\tau (\tlnn)$ decay. As an example, we show the case of $\bar{B}^0 \to D^{*+}(\to D^+ \pi^0) W^-(\to \tau^-\bar{\nu}_\tau)$, with $\tau^- \to \ell^- \bar{\nu}_\ell \nu_\tau$ decay.}
\label{fig:tautomunuquark}
\end{figure}

The specific process used to calculate the distribution is $\bar{B}^0_d \to D^{*+}(\to D^+ \pi^0) W^-(\to \tau^-\bar{\nu}_\tau)$, with $\tau^- \to \ell^- \bar{\nu}_\ell \nu_\tau$, shown in Figure~\ref{fig:decayanglestau}. The quark-level diagram for this process, with a $d$-quark spectator, is shown in Figure~\ref{fig:tautomunuquark}. The decay rates for $B$-mesons with other spectator quarks are calculated in the same way, so we skip the charge and flavor subscripts in the following. The amplitude is given by
\begin{align}
\begin{aligned}
\Amp &= \sum_{\lambda_{D^*}} \frac{G_F}{\sqrt{2}} V_{cb} \langle D \pi | D^*(\lambda_{D^*}) \rangle \langle D^*(\lambda_{D^*}) | \bar{c} \gamma_{\mu} (1-\gamma^5) b | \bar{B} \rangle (\bar{u}_\tau \gamma^{\mu} (1-\gamma^5) v_{\bar{\nu}_\tau}) \\
&\times  \frac{G_F}{\sqrt{2}}  (\bar{u}_{\nu_\tau} \gamma^{\alpha} (1-\gamma^5) u_\tau) (\bar{u}_\ell \gamma_{\alpha} (1-\gamma^5) v_{\bar{\nu}_\ell}).
\end{aligned}
\end{align}

The first line is the $\bar{B} \to D^*(\to D \pi) \tau \nu_\tau$ decay amplitude, and the second line gives the $\tlnn$ decay amplitude. We define the following quantities to simplify the amplitude:
\begin{align}\label{eq:bdlnuampdefs}
\begin{aligned}
P(\lambda_{D^*}) &\equiv \eds^{\alpha} (\lds)(p_D)_{\alpha} = \frac{1}{2 g_{D^*D\pi}} \langle D \pi | D^*(\lds)  \rangle, \\
\eds^{*\beta}(\lds) T^{V_L}_{\beta \mu} &\equiv \langle D^*(\lds) | \bar{c} \gamma_{\mu} (1-\gamma^5) b | \bar{B} \rangle,  \\
R_{V_L}^{\mu} &\equiv  (\bar{u}_{\nu_\tau} \gamma^{\alpha} (1-\gamma^5) u_\tau)  (\bar{u}_\tau \gamma^{\mu} (1-\gamma^5) v_{\bar{\nu}_\tau}) (\bar{u}_\ell \gamma_{\alpha} (1-\gamma^5) v_{\bar{\nu}_\ell}).
\end{aligned}
\end{align}
The amplitude then becomes \footnote{Note: We have made the $D^*$ and $\tau$ propagators explicit by the following replacements:
\begin{align}
\sum_{\lds} \eds^{\alpha}(\lds) \eds^{* \beta}(\lds) \rightarrow \frac{\sum_{\lds} \eds^{\alpha}(\lds) \eds^{* \beta}(\lds)}{p^2_{D^*} - M^2_{D^*} + i M_{D^*} \Gamma_{D^*}}, \quad\quad u_{\tau} \bar{u}_{\tau} \rightarrow \frac{\slashed{p}_\tau + M_\tau}{p^2_\tau - M^2_\tau + i M_\tau \Gamma_\tau}.
\end{align}}
\begin{align}
\begin{aligned}
\Amp &=  G_F^2 V_{cb} g_{D^*D\pi} \frac{1}{p^2_\tau - M^2_\tau + i M_\tau \Gamma_\tau} \frac{1}{p^2_{D^*} - M^2_{D^*} + i M_{D^*} \Gamma_{D^*}} \\
&\times \sum_{\lds} P_D(\lds) \eds^{*\beta}(\lds) T_{\beta \mu}^{V_L} R ^{\mu}_{V_L}.
\end{aligned}
\end{align}
As we can see, the hadronic part ($\eds^{*\beta}(\lds)T^{V_L}_{\beta\mu}$) and leptonic part $R^{\mu}_{V_L}$ are not Lorentz invariant quantities (because of the $\mu$ Lorentz index). Therefore, we have to use the same reference frame to express them, which is inconvenient. However, we can separate them by making use of the completeness relation of the $W$-boson polarization vectors $\ew$, which are given as follows:
\begin{align}\label{eq:completenessrelation}
\sum_{mn} \ew^{*\mu}(m) \ew^{\mu'}(n)g_{mn} = g^{\mu \mu'},
\end{align}
where $g_{mn}$ is the metric tensor and $m,n$ are the polarizations of the $W$-boson. For the $W$-boson going in negative $z$-direction in the $B$ rest frame, they are given as \cite{korner}:
\begin{align} \label{eq:W_polarization_vectors}
\begin{aligned}
\ew^{\mu} (\pm) &= \frac{1}{\sqrt{2}} (0,\pm 1, -i,0), \\
\ew^{\mu} (0) &= \frac{1}{\sqrt{q^2}} (|\vec{q}|,0,0,-q_0), \\
\ew^{\mu} (t) &= \frac{1}{\sqrt{q^2}} (q_0,0,0,-|\vec{q}|),
\end{aligned}
\end{align}
where $q_0$ and $\vec{q}$ denote the energy and momentum of the $W$-boson in the $B$ rest frame. We can now separate the hadronic and leptonic pieces as follows:
\begin{align}
\begin{aligned}
\eds^{*\beta}(\lds) T_{\beta}^{\mu, V_L} R_{\mu, V_L} &= \eds^{*\beta}(\lds) T_{\beta \mu'}^{V_L}g^{\mu'\mu} R_{\mu,V_L} \\
&=\sum_{m}  \underbrace{\eds^{*\beta}(\lds) T_{\beta \mu'}^{V_L} \ew^{*\mu'}(m)}_{H_{ V_L,m}^{\lambda_{D^*}}} g_{mm} \underbrace{\ew^{\mu}(m)R_{\mu, V_L}}_{{L}^{VA}_m},
\end{aligned}
\end{align}
where $H$ and ${L}$ are the \emph{hadronic and leptonic helicity amplitudes}, respectively. By definition, the leptonic (hadronic) helicity amplitude is simply the projection of the leptonic (hadronic) amplitude on a basis formed by the (conjugate of) $W$ polarization vectors. As we can see, both are Lorentz invariant quantities and thus can be computed in any reference frame. An additional advantage of using these quantities is that we can now write the helicity amplitudes corresponding to each NP operator $O_i$ ($i\in \{V_L, V_R, S, P, T \}$) given in Eq.~\eqref{eq:effective_hamiltonian} by simply changing the operator within the matrix element.  They are then given as follows:
\begin{align}\label{eq:lephelamp}
\begin{aligned}
{L}^{VA}_m &= {L}^{V_L}_m = {L}^{V_R}_m  \\
 &\equiv \ew^{\mu}(m) (\bar{u}_\ell \gamma_{\alpha} (1-\gamma^5) v_{\bar{\nu}_\ell}) (\bar{u}_{\nu_\tau} \gamma^{\alpha} (1-\gamma^5) \slashed{p}_\tau \gamma_{\mu} (1-\gamma^5) v_{\bar{\nu}_\tau}), \\
{L}^{SP} &= {L}^{S} = {L}^{P}  \\
 &\equiv   (\bar{u}_\ell \gamma_{\alpha} (1-\gamma^5) v_{\bar{\nu}_\ell}) (\bar{u}_{\nu_\tau} \gamma^{\alpha} (1-\gamma^5) M_\tau (1-\gamma^5) v_{\bar{\nu}_\tau}), \\
{L}^{T}_{mn} &\equiv - i  \ew^{\mu}(m) \ew^{\nu}(n) (\bar{u}_\ell \gamma_{\alpha} (1-\gamma^5) v_{\bar{\nu}_\ell})  (\bar{u}_{\nu_\tau} \gamma^{\alpha}  (1-\gamma^5) M_\tau \sigma_{\mu\nu} (1-\gamma^5) v_{\bar{\nu}_\tau}) .
\end{aligned}
\end{align}
\begin{align}\label{eq:hadhelamp}
\begin{aligned}
 H_{V_L,m}^{\lds}(q^2)
 &\equiv  \ew^{*\mu}(m) \langle D^* (p_{D^*},\eds \left(\lds)\right)| \bar c \gamma_\mu(1-\gamma^5) b| \bar B (p_B)\rangle\,,\\
 H_{V_R,m}^{\lds}(q^2)
 &\equiv  \ew^{*\mu}(m) \langle D^* (p_{D^*},\eds \left(\lds)\right)| \bar c \gamma_\mu(1+\gamma^5) b| \bar B (p_B)\rangle\,,\\
 H_{ S}^{\lds}(q^2)
 &\equiv \langle D^* (p_{D^*},\eds \left(\lds)\right)| \bar c  b| \bar B (p_B)\rangle\,,\\
 H_{ P}^{\lds}(q^2)
 &\equiv \langle D^* (p_{D^*},\eds \left(\lds)\right)| \bar c \gamma^5 b| \bar B (p_B)\rangle\,,\\
 H_{ T,mn}^{\lds}(q^2)
 &\equiv i \ew^{*\mu}(m) \ew^{*\nu}(n) \langle D^* (p_{D^*},\eds \left(\lds)\right)| \bar c \sigma_{\mu\nu}(1-\gamma^5) b| \bar B (p_B)\rangle .
\end{aligned}
\end{align}
The $i$ in the tensor helicity amplitude is a conventional choice, which also helps to make the helicity amplitude real. To compensate, a factor of $-i$ is included in the leptonic tensor amplitude.  An expression for the hadronic helicity amplitudes in terms of form factors is given in Appendix~\ref{app:tradformfactors}. We find that the scalar helicity amplitude $H_S^{\lds}$ vanishes~\cite{tensorff}, and thus, we do not write the corresponding amplitude anymore. The amplitude modulus square in terms of helicity amplitudes is written as 
\begin{align}\label{eq:bdtaunuamp}
\begin{aligned}
|\Amp|^2 &= G_F^4 |V_{cb}|^2 g_{D^*D\pi}^2 \frac{1}{\Gamma_\tau M_\tau} \pi \delta(p_\tau^2 - M_\tau^2) \frac{1}{\Gamma_{D^*} M_{D^*}} \pi \delta(p_{D^*}^2 - M_{D^*}^2)   \\
&\times \Bigg| \sum_{\lds} P(\lds) \bigg[C_S H_{ S}^{\lds} \tilde{L}^{SP} + C_{P} H_{ P}^{\lds}  \tilde{L}^{SP}  + C_{V_L} \sum_{m} g_{mm} H_{ V_L,m}^{\lds}  \tilde{L}^{VA}_m \\
&+ C_{V_R} \sum_{m} g_{mm} H_{ V_R,m}^{\lds}  \tilde{L}^{VA}_m  +C_T \sum_{m,n}  g_{mm}g_{nn} H_{{T},mn}^{\lds}  \tilde{L}^{T}_{mn} \bigg] \Bigg|^2,
\end{aligned}
\end{align}
where in the last step, we have used the narrow-width approximation on both the $\tau$ and $D^*$ propagators. The most interesting terms are the interference terms containing $C_{V_L}$. This is because NP terms, if they exist, are expected to be small ($C_{NP} \ll 1$). Squares of NP terms will be even smaller, which will be difficult to detect in experiments. However, there is a better chance that interference terms such as $C_{V_L}C_{NP}$ can be observed ($C_{V_L}=1$ for SM) in experiments, so we will focus on such terms in our analysis below.

\subsection{Phase space calculation of $\BDstaufull$ decay}\label{sec:bdtaunuphasespace}

In the process under consideration, the $\BDstaufull$ decay, the final state has six particles, of which three are neutrinos. Thus, only three particles are observable.  Usually, helicity angles are measured in the rest frame of the parent particle.  For example, the angle of the $D$-meson, $\theta_D$, is measured in the rest frame of its parent particle $D^*$. However, in the case of $\tlnn$, we cannot do that as the $\tau$ direction is not measurable in experiment. Therefore, we instead work in the $W$-rest frame, which can be determined from the hadronic side of the $B$ decay.  Thus, \emph{we work in the $W$ rest frame for the leptonic part and in the $D^*$ rest frame for the hadronic part}. 

The decay rate is given by Eq.~\eqref{eq:decayrategeneral}, where the phase space element $d\Pi_n$ is now a 6-body element.  Using recursion relations to decompose the phase space elements, we obtain
\begin{align}
\begin{aligned}
d\Pi_6 &= \frac{dq^2}{2\pi}\frac{dp_{D^*}^2}{2\pi}\frac{dp_{\tau}^2}{2\pi} \\
&\times d\Pi_2^B (B \ra D^* W) d\Pi_2^{D^*}(D^* \ra D\pi) d\Pi_2^{W}(W \ra \tau \bar{\nu}_{\tau}) d\Pi_3^{\tau}(\tlnn),
\end{aligned}
\end{align}
where $q^2 = p_W^2$, where $p_W$ is the four-momentum of the $W$-boson. Each phase space element is calculated below.

Let us consider a general decay $P_0 (p_0) \to P_1 (p_1) P_2 (p_2)$. In the $P_0$ centre of mass frame, its phase space element can be written as~\cite{pdg24}
\begin{align}
\begin{aligned}
d\Pi_2^{P} &= \int \frac{d^3 p_1 }{(2\pi)^3 2 E_1 } \frac{d^3 p_2}{ (2\pi)^3 2 E_2} (2\pi)^4 \delta^4 (p_0 - p_1 - p_2) \\
&= \frac{1}{ 4 (2\pi)^2 } \int \frac{ |p_1| d |p_1| d\Omega_1}{ M_0 }  \delta( |\vec{p}_1| - |\vec{p}_1|_{\rm root} )   ,
\end{aligned}
\end{align}
where 
\begin{align}
|\vec{p}_1|_{\rm root} = \frac{\sqrt{\lambda(M_{0}^2, M_1^2, M_2^2)}}{2 M_0},
\end{align}
where $\lambda(x,y,z) = x^2 + y^2 + z^2 - 2(xy + yz + zx)$ is the K\"allen function \cite{kallenfunction}.  Here, we have used the following property of the $\delta$ function:
\begin{align}
\delta(f(x)) = \frac{\delta(x-x_{\rm root})}{|f'(x_{\rm root})|},
\end{align}
where $x_{\rm root}$ is the root of $f(x) =0$.  Using this result, we can write the phase space elements for $W, B$ and $D^*$ decays as follows:
\\\\
\noindent
1. $\mathbf{d\Pi_2}^{W}$:

We obtain the phase space of $W \to \ell \bar{\nu}_\ell$ in the $W$ rest frame, so its four-momentum becomes $q = (\sqrt{q^2},0,0,0)$. 
\begin{align}\label{eq:wphasespace}
\begin{aligned}
d\Pi_2^W &= \int \frac{d^3p_{\ell}}{(2\pi)^3 2 E_{\ell}} \frac{d^3p_{\bar{\nu}_\ell}}{(2\pi)^3 2 E_{\bar{\nu}_\ell}} (2\pi)^4 \delta^4 (q - p_{\ell} - p_{\bar{\nu}_\ell}) \\
&= \frac{1}{4(2\pi)^2} \int \frac{|\vec{p}_\ell| d |\vec{p}_\ell| d\Omega_\ell}{\sqrt{q^2}} \delta (|\vec{p}_\ell| - \left(\frac{q^2 - M_\ell^2}{2\sqrt{q^2}} \right)),
\end{aligned}
\end{align}
where $d\Omega_\ell = d\cos\theta_\ell d\chi_\ell$, with $\theta_\ell$ and $\chi_\ell$ shown in Figure~\ref{fig:decayanglestau}.

\noindent
2. $\mathbf{d\Pi_2^{D^*}}$: 

The phase space of $D^* \to D \pi$ is calculated in the $D^*$ rest frame. 
\begin{align}\label{eq:dstarphasespace}
\begin{aligned}
d\Pi_2^{D^*} &= \int \frac{d^3 p_D }{(2\pi)^3 2 E_D } \frac{d^3 p_{\pi}}{ (2\pi)^3 2 E_{\pi}} (2\pi)^4 \delta^4 (p_{D^*} - p_{D} - p_{\pi}) \\
&= \frac{1}{ 4 (2\pi) } \int \frac{ |p_D| d |p_D| d\cos\theta_D}{  M_{D^*} }  \delta( |\vec{p}_D| - \frac{\sqrt{\lambda(M_{D^*}^2, M_D^2, M_{\pi}^2)}}{2 M_{D^*}}  )  ,
\end{aligned}
\end{align}
where in the last step, we have also integrated over the azimuthal angle $\chi_D$, and $\theta_D$ is shown in Figure~\ref{fig:decayanglestau}.

\noindent
3. $\mathbf{d\Pi_2^{B}}$:

We solve the phase space of $B \to D^* W$ in the $B$ rest frame, where $q = (q_0, \vec{q})$ is momentum of $W$. 
\begin{align}\label{eq:bphasespace}
\begin{aligned}
d\Pi_2^{B} &= \int \frac{d^3 p_{D^*}}{(2\pi)^3 2 E_{{D^*}}} \frac{d^3 q}{(2\pi)^3 2 q_0}  (2\pi)^4 \delta^4(p_B - p_{D^*} - q) \\
&=  \frac{1}{2(2\pi)}\int \frac{|\vec{p}_{D^*}|  d|\vec{p}_{D^*}| }{ M_B}  \delta(|\vec{p}_{D^*}| - \frac{\sqrt{\lambda(M_B^2, M_{D^*}^2,q^2)}}{2 M_B}) .
\end{aligned}
\end{align}

\noindent
4. $\mathbf{d\Pi_3}^{\tau}$:

The phase space for the $\tlnn$ decay can be written as
\begin{align}\label{eq:phasespacetau}
{d\Pi_3}^{\tau}= \int \frac{d^3p_{\ell}}{(2\pi)^3 2 E_{\ell}} \frac{d^3p_{\bar{\nu}_\ell}}{(2\pi)^3 2 E_{\bar{\nu}_\ell}}\frac{d^3p_{\nu_\tau}}{(2\pi)^3 2 E_{\nu_\tau}} (2\pi)^4 \delta^4(p_\tau - p_{\ell} - p_{\bar{\nu}_{\ell}} - p_{\nu_\tau}).
\end{align}
To integrate over the two neutrinos, we define the following integral for a process $P(k)\to P_1(p_1)P_2(p_2)$:
\begin{align}
I_{\alpha\beta} = \int \frac{d^3p_1}{2E_1 (2\pi)^3}\frac{d^3p_2}{2E_2 (2\pi)^3} (p_1)_{\alpha} (p_2)_{\beta} (2\pi)^4 \delta^4 (p_1 + p_2 -k)\Theta(k^2),
\end{align}
where $\Theta$ is the Heaviside step function. Usually, this function is not written explicitly, as in most cases, it does not have an impact on the final answer. However, in this phase space element, the Heaviside function will play an important role in the calculation of $\ell$-energy limits, as we will see in Section~\ref{sec:emulimits}. This integral has the general form
\begin{align}
I_{\alpha\beta} \equiv A k^2 g_{\alpha\beta} + B k_{\alpha}k_{\beta},
\end{align}
which allows us to define two Lorentz invariant quantities
\begin{align}\label{eq:intliquantites}
I_{\alpha\beta} g^{\alpha\beta} = 4Ak^2 + Bk^2, \quad\quad I_{\alpha\beta}k^{\alpha}k^{\beta} = A k^4 + B k^4.
\end{align}
Let us now evaluate these quantities. We choose the centre of mass frame of particles $P_1$ and $P_2$ (which we consider to be massless). Therefore, 
\begin{align}
p_1^{\mu} = (|\vec{p_1}|,\vec{p_1}), \quad\quad p_2^{\mu} = (|\vec{p_1}|,-\vec{p_1}), \quad\quad k = (E_k,0,0,0).
\end{align}
The first integral is
\begin{align}\label{eq:int1}
\begin{aligned}
I_{\alpha\beta} g^{\alpha\beta} &= \int \frac{d^3p_1}{2E_1 (2\pi)^3}\frac{d^3p_2}{2E_2 (2\pi)^3} (p_1 . p_2) (2\pi)^4 \delta^4 (p_1 + p_2 -k)\Theta(k^2) \\
&=\frac{1}{(4\pi)^2} \pi E_k^2,
\end{aligned}
\end{align}
and the second integral is 
\begin{align}\label{eq:int2}
\begin{aligned}
I_{\alpha\beta}k^{\alpha}k^{\beta} &= \int \frac{d^3p_1}{2E_1 (2\pi)^3}\frac{d^3p_2}{2E_2 (2\pi)^3} (p_1 . p_2)^2 (2\pi)^4 \delta^4 (p_1 + p_2 -k)\Theta(k^2) \\
&=\frac{1}{(4\pi)^2} \frac{1}{2}\pi E_k^4.
\end{aligned}
\end{align}
Now inserting Eq.~\eqref{eq:int1} and Eq.~\eqref{eq:int2} in Eq.~\eqref{eq:intliquantites}, we can solve the two equations to obtain (using $k^2 = E_k^2$) $A = \frac{1}{(4\pi)^2}\frac{\pi}{6}$ and $B = \frac{1}{(4\pi)^2}\frac{\pi}{3}$. Therefore, the integral is 
\begin{align}
\begin{aligned}
I_{\alpha\beta} &= \int \frac{d^3p_1}{2E_1 (2\pi)^3}\frac{d^3p_2}{2E_2 (2\pi)^3} (p_1)_{\alpha} (p_2)_{\beta} (2\pi)^4 \delta^4 (p_1 + p_2 -k)\Theta(k^2) \\
&= \frac{1}{(4\pi)^2}\frac{\pi}{6}(k^2 g_{\alpha\beta} + 2 k_{\alpha}k_{\beta}).
\end{aligned}
\end{align}
Now to use this result in Eq.~\eqref{eq:phasespacetau}, we identify $P_1$ and $P_2$ as $\nu_\tau$ and $\bar{\nu}_\ell$, respectively (changing $1 \leftrightarrow 2$ will lead to the same result). By doing this, we find
\begin{align}
\begin{aligned}
&\int \frac{d^3p_{\bar{\nu}_\ell}}{2E_{\bar{\nu}_\ell} (2\pi)^3}\frac{d^3p_{\nu_\tau}}{2E_{\nu_\tau} (2\pi)^3} (p_{\bar{\nu}_\ell})_{\alpha} (p_{\nu_\tau})_{\beta} (2\pi)^4 \delta^4 ((p_\tau - p_\ell)-p_{\bar{\nu}_\ell} - p_{\nu_\tau} ) \Theta((p_\tau - p_\ell)^2) \\
&= \frac{1}{(4\pi)^2}\frac{\pi}{6}[(p_\tau - p_\ell)^2 g_{\alpha\beta} + 2 (p_\tau - p_\ell)_{\alpha} (p_\tau - p_\ell)_{\beta}].
\end{aligned}
\end{align}
Therefore, we obtain
\begin{align}\label{eq:tauphasespace}
\begin{aligned}
d\Pi_3^\tau p_{\bar{\nu}_\ell}^{\sigma} p_{\nu_{\tau}}^{\delta} = \frac{1}{96 (2\pi)^4} &\int {\sqrt{E_\ell^2 - M_\ell^2} dE_\ell d\Omega_{\ell}} \Theta((p_\tau - p_\ell)^2)  \\
&\times \left[ (p_\tau - p_\ell)^2 g^{\sigma \delta} + 2 (p_\tau - p_\ell)^{\sigma} (p_\tau - p_\ell)^{\delta} \right]  ,
\end{aligned}
\end{align}
with $d\Omega_\ell = d\!\cos\theta_\ell d\chi_\ell$. The calculation of the limits of $E_\ell$ is non-trivial in the $W$ rest frame, as discussed below.

\subsection{Kinematical range of $E_\ell$ in $\tlnn$ decay in the $W$ rest frame}\label{sec:emulimits}
The calculation of the kinematical range of $E_\ell$ is a bit subtle in this case, as the limits depend upon other variables.  Recalling the step function $\Theta((p_\tau - p_\ell)^2)$ that appears in Eq.~\eqref{eq:tauphasespace}, in order to have a nonzero value of this function,  the following condition must be satisfied:
\begin{align}\label{eq:taumuinequality}
(p_\tau - p_\ell)^2 > 0.
\end{align}
In the $W$ rest frame, let us define the angle between the $\tau$ and $\ell$ by $\theta_{\tau\ell}$.  The above inequality then gives
\begin{align}\label{eq:cosinequality}
\cos\theta_{\tau\ell} > \frac{ 2 E_\tau E_\ell - M_\tau^2 - M_\ell^2 }{2|\vec{p}_\tau| \sqrt{E_\ell^2-M_\ell^2}} ,
\end{align}
where all the energies and momenta are in the $W$ rest frame, and the value of $|\vec{p}_\tau|$ is given in Eq.~\eqref{eq:momentummagnitude}, and $E_\tau = \sqrt{|\vec{p}_\tau|^2 + M_\tau^2}$. Since $1\geq \cos\theta_{\tau\ell} \geq -1$, we obtain,
\begin{align}\label{eq:emuinequality}
 \frac{ 2 E_\tau E_\ell - M_\tau^2 - M_\ell^2 }{2|\vec{p}_\tau| \sqrt{E_\ell^2-M_\ell^2}}  \geq -1 \Leftrightarrow E_\ell \geq \frac{M_\tau^4 + M_\ell^2 q^2}{2 M_\tau^2 \sqrt{q^2}}.
\end{align}
Whenever the above inequality is satisfied, the lower limit of $\cos\theta_{\tau\ell}$ will be given by the expression in Eq.~\eqref{eq:cosinequality}. Conversely, whenever the inequality is not satisfied, the lower limit of $\cos\theta_{\tau\ell}$ is $-1$.
\par
\begin{figure}
\centering
\includegraphics[scale=0.5]{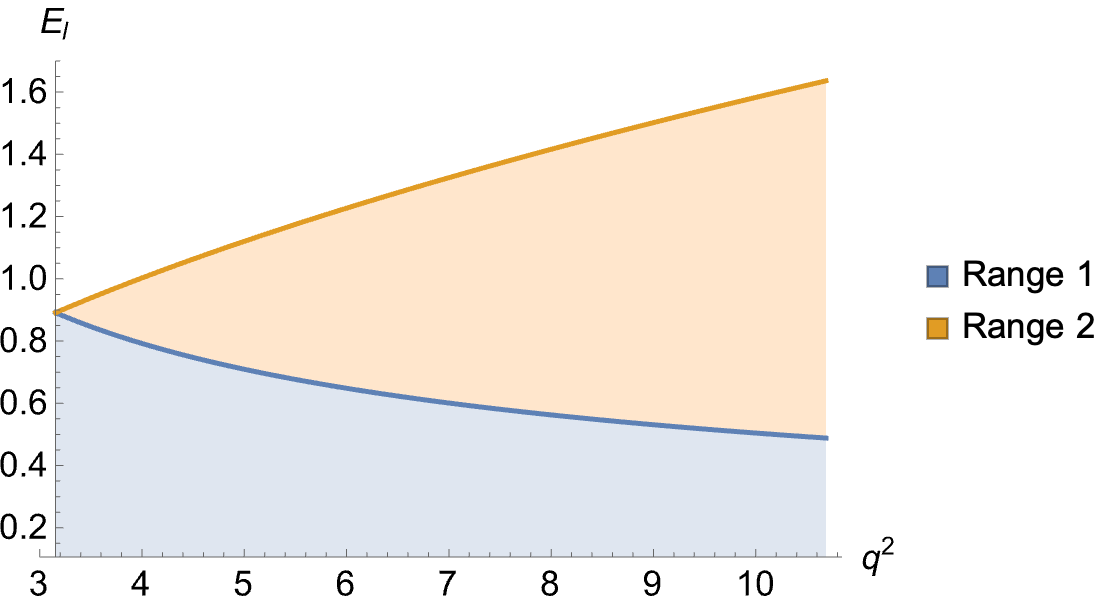}
\caption{\small The two regions of phase space integration in $q^2$ and $E_\ell$. The blue region corresponds to Range 1, and the orange region corresponds to Range 2, where Range 1 and Range 2 are given in Table~\ref{table:emulimits}.}
\label{fig:emulimits}
\end{figure}

 
The next step is to find the upper and lower limits of $E_\ell$. The lower limit can be simply inferred by putting $\ell$ at rest, giving $E_{\ell, {\rm min}} = M_\ell$. The upper limit can be found by setting $\cos\theta_{\tau\ell}=1$ in Eq.~\eqref{eq:taumuinequality}, which gives $E_{\ell, {\rm max}} = \frac{q^2 + M_\ell^2}{2\sqrt{q^2}}$. Using these limits and that from Eq.~\eqref{eq:emuinequality}, the range of integration can be split into two parts \cite{emulimitschina}. The limits are given in Table~\ref{table:emulimits}, and the corresponding phase space integration region is shown in Figure~\ref{fig:emulimits}.
\par
Finally, we note that the limits in the Table~\ref{table:emulimits} are given in terms of $\cos\theta_{\tau\ell}$, while the differential decay rate in Eq.~\eqref{eq:angdisttauv1} is given in terms of $\cos\theta_\tau$ (the angle between the $\tau$ and the opposite of the $B$ direction in the $W$ rest frame, which we define as the $z$-axis.) However, given that both definitions are in the same reference frame, we can switch between the two with a simple rotation of coordinate axes. In addition, since the phase space integral that we measure in Eq.~\eqref{eq:wphasespace} is rotationally invariant, we can simply work in a frame in which the z-axis is rotated along the $\ell$ direction and use the limits given in Table~\ref{table:emulimits}.
\begin{table}
\begin{center}
\renewcommand{\arraystretch}{1.5}
\begin{tabular}{|c||c|c|c|c|}
\hline
Range & $E_\ell$ limits & $ \cos\theta_{\tau\ell}$ limits \\
\hline 
Range 1 (R1) & $M_\ell \leq E_\ell \leq \frac{M_\tau^4 + M_\ell^2 q^2}{2 M_\tau^2 \sqrt{q^2}}$ & $-1 \leq \cos\theta_{\tau\ell} \leq 1$ \\
\hline 
Range 2 (R2) & $\frac{M_\tau^4 + M_\ell^2 q^2}{2 M_\tau^2 \sqrt{q^2}} \leq E_\ell  \leq  \frac{q^2 + M_\ell^2}{2\sqrt{q^2}}$ & $ \frac{ 2 E_\tau E_\ell - M_\tau^2 - M_\ell^2 }{2|\vec{p}_\tau| \sqrt{E_\ell^2-M_\ell^2}}  \leq \cos\theta_{\tau\ell} \leq 1$ \\
 \hline
\end{tabular}
\end{center}
\caption{\small The two ranges of integration, with $E_\ell$ and $\cos\theta_{\tau\ell}$ limits in the $W$ rest frame. }
\label{table:emulimits}
\end{table}


\subsection{Angular distribution expression}\label{sec:angdistexp}
Putting together the phase space elements from Eqs.~\eqref{eq:wphasespace},~\eqref{eq:dstarphasespace},~\eqref{eq:bphasespace} and ~\eqref{eq:tauphasespace}, and the amplitude given in Eq.~\eqref{eq:bdtaunuamp} in the decay rate formula (Eq.~\eqref{eq:decayrategeneral}), we obtain
\begin{align}\label{eq:angdisttauv1}
\begin{aligned}
&\frac{d\Gamma}{dq^2 dE_\ell d\Omega_{\ell} d\cos\theta_D d\Omega_\tau} \\
&= \frac{3}{32(4\pi)^5} \frac{G_F^2 |V_{cb}|^2 |\eta_{EW}|^2\  \mathcal{B}(D^* \ra D \pi)  \mathcal{B}(\tlnn) E_\ell |\vec{p}_{D^*}|   |\vec{p}_\tau| }{|\vec{p}_{D}|^2 M_B^2 M_\tau^6 \sqrt{q^2} } \\
&\times \sum_{\lambda_{\ell, \bar{\nu}_\ell},\nu_\tau,\bar{\nu}_{\tau}} |\Amp_{V_L}+\Amp_{V_R}+\Amp_{P}+\Amp_{T}|^2,
\end{aligned}
\end{align}
where
\begin{align}\label{eq:momentummagnitude}
|\vec{p}_{D^*}| &= \frac{\sqrt{\lambda(M_B^2, M_{D^*}^2,q^2)}}{2 M_B},  \quad |\vec{p}_D| &= \frac{\sqrt{\lambda(M_{D^*}^2, M_D^2, M_{\pi}^2)}}{2 M_{D^*}}, \quad
|\vec{p}_\tau| &= \frac{q^2 - M_\tau^2}{2\sqrt{q^2}}.
\end{align}
The amplitudes $\Amp_{V_L,V_R,P,T}$ are given as follows:
\begin{align}\label{eq:npamps}
\Amp_{V_L} &=  \, C_{V_L} \sum_{\lambda_{D^*}= \pm,0} P(\lambda_{D^*}) \sum_{\lambda= t,\pm,0} g_{\lambda\lambda} H^{\lambda_{D^*}}_{\rm V_L,\lambda} L_{\lambda}^{\rm VA}, \\
\Amp_{V_R} &=  \, C_{V_R} \sum_{\lambda_{D^*}= \pm,0} P(\lambda_{D^*}) \sum_{\lambda= t,\pm,0} g_{\lambda\lambda} H^{\lambda_{D^*}}_{\rm V_R,\lambda} L_{\lambda}^{\rm VA}, \\
\Amp_{\rm P} &=  \, C_{P} \sum_{\lambda_{D^*}= \pm,0} P(\lambda_{D^*}) H^{\lambda_{D^*}}_{P} L^{\rm SP}, \\
\Amp_{ T} &=  \, C_{ T}
\sum_{\lambda_{D^*}= \pm,0} P(\lambda_{D^*})
\sum_{\lambda= t,\pm,0} \sum_{\lambda'= t,\pm,0}g_{\lambda\lambda} g_{\lambda'\lambda'}
H^{\lambda_{D^*}}_{\rm T,\lambda \lambda'} L^{T}_{\lambda \lambda'}.
\end{align}
The SM branching fraction expressions of $\Dsdpi$ and $\tlnn$ used in Eq.~\eqref{eq:angdisttauv1} are given as follows:
\begin{align}
\mathcal{B}(D^* \ra D \pi) = \frac{g_{D^*D\pi}^2 |\vec{p}_D|^3}{6 \pi M_{D^*}^2\Gamma_{D^*}}, \quad \quad \mathcal{B}(\tlnn) = \frac{G_F^2 M_\tau^5}{192 \pi^3 \Gamma_\tau}.
\end{align}
In Eq.~\eqref{eq:angdisttauv1}, the helicity angles of the $\tau$ still appear, which are not measurable in experiment. Therefore, to write the measurable angular distribution, we must integrate out these angles after inserting the amplitudes from Eq.~\eqref{eq:npamps} in Eq.~\eqref{eq:angdisttauv1}.  However, as explained in Section~\ref{sec:emulimits}, there are two ranges of integration (R1 and R2) of $\cos\theta_\tau$ and $E_\ell$. To obtain the full distribution, we need to integrate over the two ranges separately and add the results at the end. The final measurable angular distribution is then written as follows: 
\begin{align}\label{eq:rateJtau}
\begin{aligned}
\frac{d \Gamma^{\rm r}( \bar{B} \rightarrow D^{*} (\rightarrow D \pi) \tau^- (\to \ell \bar{\nu}_\ell \nu_\tau)\bar \nu_\tau)}{d w d E_{\ell} d\!\cos\theta_D d\!\cos\theta_{\ell} d \chi_{\ell}} 
&= \frac{3 G_F^2 \left|V_{cb}\right|^2 |\eta_{\rm EW}|^2 M_{D^*} \mathcal{B}(D^{*} \to D \pi) \mathcal{B}(\tlnn) }{16(4\pi)^5 M_B^2 M_\tau^6 |\vec{p}_D|^2}   \\
&\times \frac{|\vec{p}_{D^*} (w)| |\vec{p}_\tau (w)| E_{\ell}}{\sqrt{1+r^2-2wr}}  \Big\{ J^{\rm r}_{1s} \sin^2\theta_D+J^{\rm r}_{1c}\cos^2\theta_D \\
&+(J^{\rm r}_{2s} \sin^2\theta_D+J^{\rm r}_{2c}\cos^2\theta_D )\cos 2\theta_\ell   \\
& +J^{\rm r}_3 \sin^2\theta_D\sin^2\theta_\ell\cos 2\chi_\ell  \\
&  +J^{\rm r}_4\sin 2\theta_D\sin 2\theta_\ell \cos\chi_\ell 
+J^{\rm r}_5 \sin 2\theta_D\sin\theta_\ell\cos\chi_\ell \\ 
& +(J^{\rm r}_{6s} \sin^2\theta_ D+J^{\rm r}_{6c}\cos^2\theta_D)\cos\theta_\ell  \\
& +J^{\rm r}_7 \sin 2\theta_D\sin\theta_\ell \sin\chi_\ell+J^{\rm r}_8\sin 2\theta_D \sin 2\theta_\ell\sin\chi_\ell  \\
& +J^{\rm r}_9 \sin^2\theta_D\sin^2\theta_\ell \sin2\chi_\ell  \Big\},
\end{aligned}
\end{align}
where $r=M_{D^*}/M_B$, and we have changed the variable from $q^2$ to $w$ using the following relation:
\begin{align}\label{eq:qwrelation}
w = v_B.v_{D^*}= \frac{M_B^2 + M_{D^*}^2-q^2}{2 M_B M_{D^*}}.
\end{align}
We have put the superscript "r" over $\Gamma$ and $J$ to keep track of the range of integration to use, R1 and R2.  That is to say, when r = R1(R2),  it implies that the above decay rate has been obtained by integrating $\cos\theta_\tau$ using the limits given in the first(second) row of Table~\ref{table:emulimits}, and while integrating over $E_\ell$ in subsequent steps, we should use the limits given in the first(second) row of Table~\ref{table:emulimits}.
This leads to two sets of $J$-functions, $J^{\rm R1}_i$ and $J^{\rm R2}_i$, with $i \in \{1s,1c,2s,2c,3,4,5,6s,6c,7,8,9 \}$. These $J$ functions are functions of $E_\ell$ and $q^2$ (or $w$). Since these expressions are very long, they are given in a Mathematica file, available on the following link:
\par
\href{https://github.com/tejhaskapoor/Angular-decay-distribution-angular-functions-of-B-to-D-tau-nu-decay-with-tau-to-mu-nu-nu.git}{Link to Mathematica file}
\par
Here, we give a table showing which combination of NP Wilson coefficients appear in which $J$-function.
\begin{table}[H]
\begin{center}
\renewcommand{\arraystretch}{1.4}
\begin{tabular}{|c||c|c|c|c|}
\hline
J function & LH & LH-RH & LH-PS & LH-T  \\
\hline 
$J_{1s}$ & $|C_{V_L}^2|$ & ${\rm Re}(C_{V_L}C_{V_R}^*)$ & 0 & ${\rm Re}(C_{V_L}C_{T}^*)$ \\
\hline 
$J_{1c}$ & $|C_{V_L}^2|$ & ${\rm Re}(C_{V_L}C_{V_R}^*)$ & ${\rm Re}(C_{V_L}C_{P}^*)$ & ${\rm Re}(C_{V_L}C_{T}^*)$ \\
\hline 
$J_{2s}$ & $|C_{V_L}^2|$ & ${\rm Re}(C_{V_L}C_{V_R}^*)$ & 0 & ${\rm Re}(C_{V_L}C_{T}^*)$ \\
\hline 
$J_{2c}$ & $|C_{V_L}^2|$ & ${\rm Re}(C_{V_L}C_{V_R}^*)$ & 0 & ${\rm Re}(C_{V_L}C_{T}^*)$ \\
\hline 
$J_3$ & $|C_{V_L}^2|$ & ${\rm Re}(C_{V_L}C_{V_R}^*)$ & 0 & ${\rm Re}(C_{V_L}C_{T}^*)$ \\
\hline 
$J_4$ & $|C_{V_L}^2|$ & ${\rm Re}(C_{V_L}C_{V_R}^*)$ & 0 & ${\rm Re}(C_{V_L}C_{T}^*)$ \\
\hline 
$J_5$ & $|C_{V_L}^2|$ & ${\rm Re}(C_{V_L}C_{V_R}^*)$ & ${\rm Re}(C_{V_L}C_{P}^*)$ & ${\rm Re}(C_{V_L}C_{T}^*)$ \\
\hline 
$J_{6s}$ & $|C_{V_L}^2|$ & 0 & 0 & ${\rm Re}(C_{V_L}C_{T}^*)$ \\
\hline 
$J_{6c}$ & $|C_{V_L}^2|$ & ${\rm Re}(C_{V_L}C_{V_R}^*)$ & ${\rm Re}(C_{V_L}C_{P}^*)$ & ${\rm Re}(C_{V_L}C_{T}^*)$ \\
\hline 
$J_7$ & 0 & ${\rm Im}(C_{V_L}C_{V_R}^*)$ & ${\rm Im}(C_{V_L}C_{P}^*)$ & ${\rm Im}(C_{V_L}C_{T}^*)$ \\
\hline 
$J_8$ & 0 & ${\rm Im}(C_{V_L}C_{V_R}^*)$ & 0 & ${\rm Im}(C_{V_L}C_{T}^*)$ \\
\hline 
$J_9$ & 0 & ${\rm Im}(C_{V_L}C_{V_R}^*)$ & 0 & ${\rm Im}(C_{V_L}C_{T}^*)$ \\
\hline 
\end{tabular}
\end{center}
\caption{\small $J$-function dependence on NP Wilson coefficients. }
\label{tab:jfunctioncombinations}
\end{table}


\section{Unbinned angular analysis of $\BDstaufull$ with simulated data}\label{sec:unbinnedanalysis}

In this section, we perform an unbinned analysis of $\BDstaufull$ to investigate the sensitivity to NP parameters. In our calculations, we will neglect the mass of the muon, $\mu$, thus treating $e$ and $\mu$ on the same footing. At the end, we will justify this approximation by discussing the impact of adding the $\mu$ mass in the calculation. 
\par
Since there is no actual experimental data available for the angular distribution of $\BDstaufull$ decay, to demonstrate the analysis, we will generate simulated data for the angular distribution. To generate this data, we use all the fitted parameters (i.e. form factors and $V_{cb}$) measured by Belle~\cite{bellefittechnique} from $\BDslnudpi$ decay case as truth values\footnote{Since the Belle results do not give the value of $\mathcal{F}_2$, we take its value from Fermilab Lattice data~\cite{fermilab}.}, and put them in the distribution obtained in Section~\ref{sec:angdistfullbdtaunu} to obtain simulated data for the angular distribution of $\BDstaufull$ decay. The method to generate simulated data is described in Appendix~\ref{app:statprocedure}. During the fit, we will add lattice QCD data for form factors \cite{fermilab}\cite{jlqcd} and constrain $|V_{cb}|$ and $\BDstaunu$ branching ratio by their global averages.
We disregard the central values of NP parameters, as they are strongly constrained by the simulated data. Thus, deviations of NP coefficients from zero should not be considered as signals for NP. However, the statistical errors and the correlations between the measurements that we obtain using this method are reliable estimates of future sensitivities.
\par
\begin{figure}
\centering
\begin{subfigure}{1\textwidth}
\centering
\includegraphics[width=\textwidth]{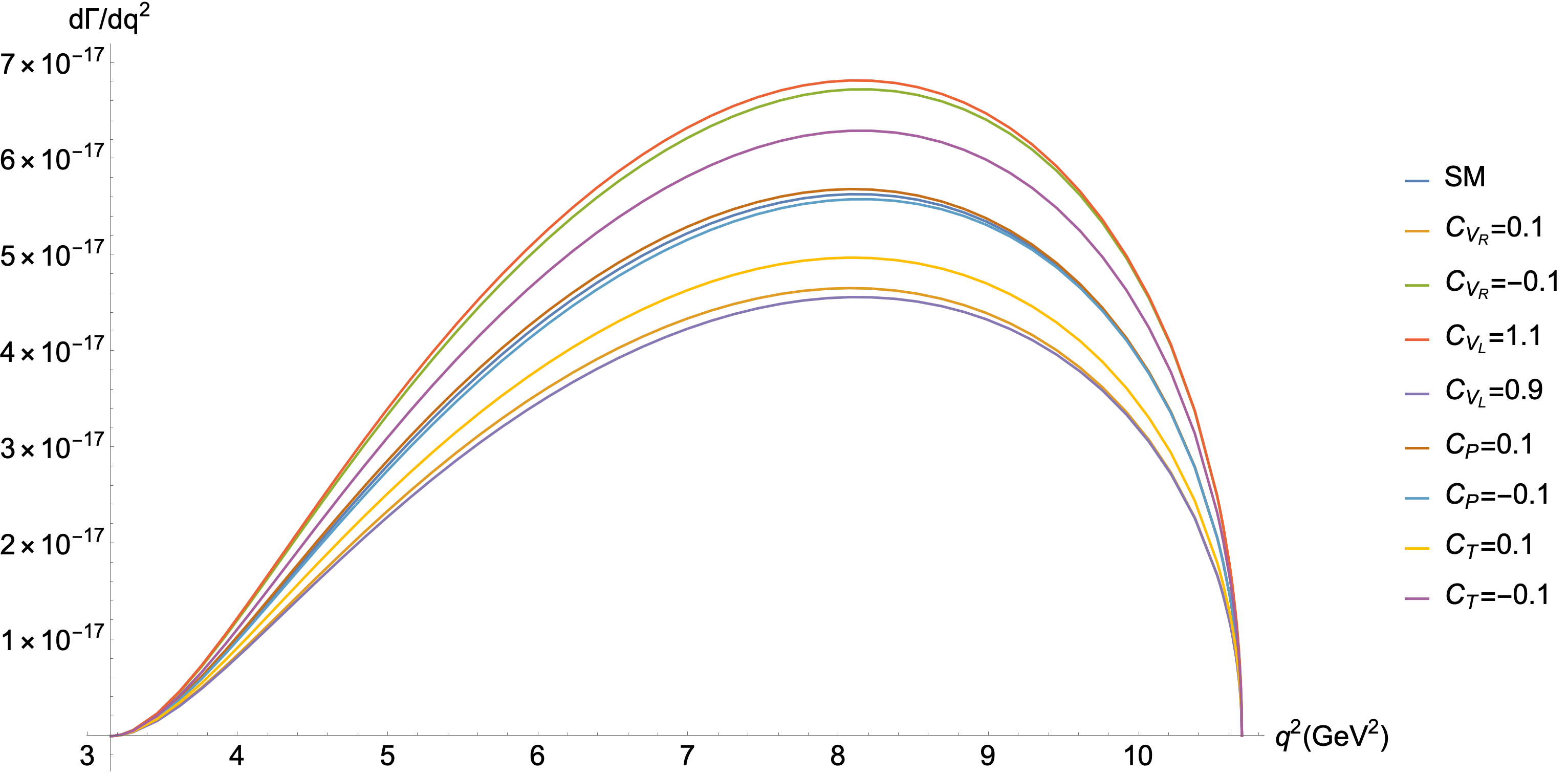}
\end{subfigure}
\hfill
\begin{subfigure}{1\textwidth}
\centering
\includegraphics[width=\textwidth]{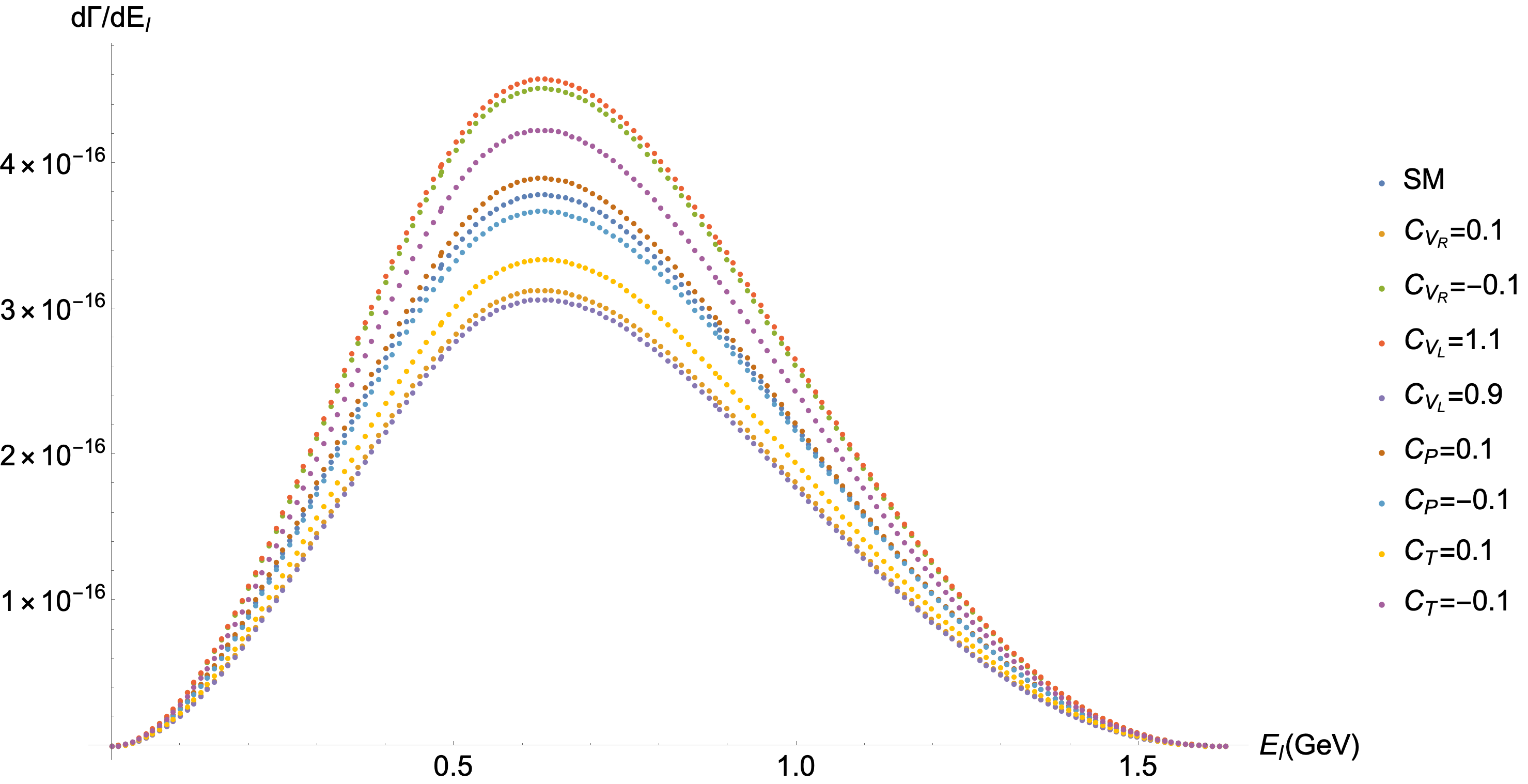}
\end{subfigure}
\caption{\small Decay rates as a function of $q^2$ and $E_\ell$ for different NP scenarios. On normalizing with the decay rates, the plot shapes are very similar. Therefore, such one dimensional analyses are insensitive to NP.}
\label{fig:q2emuplots}
\end{figure}

Before moving to the unbinned analysis, let us examine the information we can obtain through a binned analysis.  For this, we checked the decay rates with only one visible kinematic variable, for example $d\Gamma/dx$, where $x$ can be $q^2, E_\ell, \cos\theta_D,\cos\theta_\ell$ or $\chi_\ell$. However, we found that integration over the remaining kinematic variables resulted in a drastic loss of information, and these observables were barely sensitive to different NP scenarios. For illustration, we show the plots of $d\Gamma/dq^2$ and  $d\Gamma/dE_\ell$ in Figure~\ref{fig:q2emuplots}. We found that when normalized to the fully integrated decay rate, the plots were not easily distinguishable for reasonable values of NP Wilson coefficients (i.e. of $O(0.1)$). It would be impossible to compare it to experimental data, as form factor error effects would mask any NP signature. Therefore, we proceed directly to the unbinned analysis. 

\subsection{Unbinned maximum likelihood method}

Let us now illustrate how to perform the unbinned fit with the maximum likelihood method. While the method is similar to that of~\cite{emizh,bdlnumypaper} for $\BDslnudpi$ case,  we need to take care of the fact that we now have two sets of $J$-functions, corresponding to the two ranges of integration of $\ell$ energy. 
\par 
We start by obtaining the normalized Probability Density Function (PDF) in terms of the $11\times2$ independent angular observables $J_i^{\rm R1}$ and $J_i^{\rm R2}$. For this, we first give the expression of the decay rate (Eq.~\eqref{eq:rateJtau}) integrated over all the angles
\begin{align}
\begin{aligned}
\frac{d\Gamma^{\rm r}}{dw dE_\ell} &= \frac{3}{16(4\pi)^5} \frac{G_F^2 \left|V_{cb}\right|^2 |\eta_{\rm EW}|^2 M_{D^*} \mathcal{B}(D^{*} \to D \pi) \mathcal{B}(\tlnn) }{M_B^2 M_\tau^6 |\vec{p}_D|^2}  \\
&\times \frac{|\vec{p}_{D^*} (w)| |\vec{p}_\tau (w)| E_{\ell}}{\sqrt{1+r^2-2wr}} 
 \frac{8\pi}{9} \left( 6J^{\rm r}_{1s} + 3J^{\rm r}_{1c} - 2J^{\rm r}_{2s} - J^{\rm r}_{2c} \right)
 \end{aligned}
\end{align}
where r is R1 or R2 corresponding to the two integration ranges of $E_\ell$ as defined in Table~\ref{table:emulimits}.  As usual, the total decay rate is the sum of both decay rates.  
\par 
At this point, let us define the following quantities 
\begin{align}\label{eq:jfuncintegration}
{J'}_i(w) &= \frac{|\vec{p}_{D^*} (w)| |\vec{p}_\tau (w)| E_{\ell}}{\sqrt{1+r^2-2wr}} \int_{M_\ell}^{\frac{M_\tau^4 + M_\ell^2 q^2(w)}{2 M_\tau^2 \sqrt{q^2(w)}}} E_\ell J_i^{\rm R1} (w,  E_\ell) dE_\ell \\
{J''}_i(w) &= \frac{|\vec{p}_{D^*} (w)| |\vec{p}_\tau (w)| E_{\ell}}{\sqrt{1+r^2-2wr}} \int_{\frac{M_\tau^4 + M_\ell^2 q^2(w)}{2 M_\tau^2 \sqrt{q^2(w)}}}^{\frac{q^2(w) + M_\ell^2}{2\sqrt{q^2(w)}}} E_\ell J_i^{\rm R2} (w,  E_\ell) dE_\ell
\end{align}
where we use Eq.~\eqref{eq:qwrelation} in the integration limits. In these equations, we have collected all the $w$-dependent quantities and integrated the $J$-functions over the two ranges of $E_\ell$. We can add the above two quantities (as the integration range of $w$ is common for both of them) to define
\begin{align}
\tilde{J}_i(w) = {J'}_i(w) + {J''}_i(w).
\end{align}
The total decay rate as a function of $w$ is
\begin{align}
\begin{aligned}
\frac{d\Gamma}{dw} &=  \frac{3}{16(4\pi)^5} \frac{G_F^2 \left|V_{cb}\right|^2 |\eta_{\rm EW}|^2 M_{D^*} \mathcal{B}(D^{*} \to D \pi) \mathcal{B}(\tlnn) }{M_B^2 M_\tau^6 |\vec{p}_D|^2} \\
& \times \frac{8 \pi}{9} \left( 6 \tilde{J}_{1s}(w) + 3 \tilde{J}_{1c}(w) - 2 \tilde{J}_{2s}(w)- \tilde{J}_{2c}(w) \right)
\end{aligned}
\end{align}
Then, we divide the total $w$ range of integration into 10 bins and integrate over each $w$-bin to define the integrated decay rate as 
\begin{align}\label{eq:decayratewbintau}
\begin{aligned}
\langle \Gamma \rangle_{w-\rm bin} &= \frac{3}{16(4\pi)^5} \frac{G_F^2 \left|V_{cb}\right|^2 |\eta_{\rm EW}|^2 M_{D^*} \mathcal{B}(D^{*} \to D \pi) \mathcal{B}(\tlnn) }{M_B^2 M_\tau^6 |\vec{p}_D|^2} \\
&\times \frac{8\pi}{9} \left( 6 \langle \tilde{J}_{1s} \rangle_{ w-\rm bin} + 3  \langle \tilde{J}_{1c} \rangle_{w-\rm bin} - 2 \langle \tilde{J}_{2s} \rangle_{ w-\rm bin} - \langle \tilde{J}_{2c} \rangle_{w-\rm bin} \right)
 \end{aligned}
\end{align}
where 
\begin{align}
\langle 
\tilde{J}_i \rangle_{w-{\rm bin}}\equiv \int_{w-{\rm bin}} \tilde{J}_i(w) dw   
\end{align}
Hereafter, the index "$w$-bin" is implicit. The normalized PDF {for each $w$-bin} now can be {written} by the new normalized angular coefficients {$\vec{\langle g\rangle}=\langle g_i\rangle$} as
\begin{align}\label{eq:normpdftau}
\begin{aligned}
\hat{f}_{\langle \vec{g} \rangle} (\cos\theta_D, \cos\theta_\ell,\chi_\ell) 
&= \frac{9}{8\pi} 
 \Big\{ \frac{1}{6}(1 - 3 \langle g_{1c} \rangle + 2 \langle g_{2s} \rangle +\langle g_{2c} \rangle) \sin^2\theta_D + \langle g_{1c} \rangle \cos^2 \theta_D  \\
 &+ (\langle g_{2s} \rangle \sin^2 \theta_D  + \langle g_{2c} \rangle \cos^2 \theta_D )\cos 2\theta_\ell  \\
 &+ \langle g_{3} \rangle \sin^2 \theta_D \sin^2 \theta_\ell  \cos 2\chi_\ell  
 + \langle g_{4} \rangle \sin 2\theta_D \sin 2\theta_\ell \cos \chi_\ell  \\
 &+ \langle g_{5} \rangle \sin 2\theta_D \sin \theta_\ell  \cos \chi_\ell  
 + (\langle g_{6s} \rangle \sin^2 \theta_D  + \langle g_{6c} \rangle \cos^2 \theta_D ) \cos \theta_\ell  \\
 &+ \langle g_{7} \rangle \sin 2\theta_D  \sin \theta_\ell  \sin \chi_\ell  
 + \langle g_8 \rangle \sin 2\theta_D \sin 2\theta_\ell \sin \chi_\ell  \\
 &+ \langle g_{9} \rangle \sin^2 \theta_D  \sin^2 \theta_\ell  \sin 2\chi_\ell  \Big\}
\end{aligned}
\end{align}
where
\begin{equation} \label{eq:gifunctau}
\langle g_i \rangle \equiv \frac{\langle \tilde{J}_i \rangle}{6 \langle \tilde{J}_{1s} \rangle + 3 \langle \tilde{J}_{1c} \rangle - 2\langle \tilde{J}_{2s} \rangle - \langle \tilde{J}_{2c} \rangle}
\end{equation}
We can then use the log-likelihood to determine the angular coefficients $\langle \vec{g}\rangle $ from the data: 
\begin{equation}\label{eq:likelihoodtau}
\mathcal{L}(\langle \vec{g} \rangle) = \sum_{i=1}^{N} \ln \hat{f}_{\langle \vec{g} \rangle}(e_i)
\end{equation}
where $e_i$ denotes the experimental events in ($\cos\theta_D, \cos\theta_\ell,\chi_\ell$) and $N$ are the number of events for each $w$-bin. 
The best-fit value for $\langle g_i \rangle$ is obtained by maximizing the likelihood 
\begin{align}
\left. \frac{\partial \mathcal{L}}{\partial \langle g_i \rangle}\right|_{\langle g_i \rangle=\langle g_i \rangle_{\rm fit}} =0
\end{align}
The inverse of the covariance matrix is obtained as
\begin{equation}\label{eq:covmatinversetau}
V^{-1}_{ij}=-\left.\frac{\partial^2 \mathcal{L}}{\partial \langle g_i \rangle\partial \langle g_j \rangle}\right|_{\langle g_{i} \rangle=\langle g_{i} \rangle_{\rm fit}}
\end{equation}
Eq.~\eqref{eq:covmatinversetau} provides the 11x11 matrix for each $w$-bin. There are 10 such matrices since we have 10 bins in $w$.  The method to generate covariance matrices for a simple sensitivity study is described in Appendix~\ref{app:statprocedure}. \cn{One of the advantages of this method is that the covariance matrix obtained is not sensitive to slight changes in the values of $\langle g_i \rangle_{\rm fit}$, and thus, to the input form factors. Hence, it is a robust method to evaluate the sensitivities and correlations of form factors and NP parameters. }

\subsection{Fitting the theory parameters with the simulated data together with the lattice inputs}
The procedure followed for a $\chi^2$ fit is similar to that for the $\BDslnudpi$ decay, as done in~\cite{bdlnumypaper}.
Instead of performing a Toy Monte Carlo with generated events, we obtain directly $\langle g_i \rangle_{\rm fit}$ as well as $V_{ij}$ from the normalized PDF $\hat{f}_{\langle \vec{g} \rangle}$ with \emph{truth} inputs (i.e. the fitted form factors and $V_{cb}$ in~\cite{belle19} as mentioned earlier). Note that the the $\langle \vec{g} \rangle_{\rm fit}$ is obtained from $\BDslnu$ data, not $\BDstaunu$ data. Thus, the central values of NP parameters obtained from this data are not reliable, but the statistical errors and correlations are trustworthy.  The detailed procedure is given in the Appendix~\ref{app:statprocedure}. 
\par 
Using the generated $\langle g_i \rangle_{\rm fit}$ and $V_{ij}$ values, we can fit the theory parameters to the ’experimentally’ obtained angular coefficients
and their correlations by using the following $\chi^2$:
\begin{align}\label{eq:chi2angle}
&\chi^2_{\rm angle}(\vec{a}_{\rm BGL}, C_{\rm NP}) = \\
&\sum_{w-{\rm bin} = 1}^{10} \left[ N_{w-{\rm bin}}\hat{V}^{-1}_{ij} \left( \langle g_i^{\rm exp}  \rangle - \langle g_i^{\rm th} (\vec{a}_{\rm BGL}, C_{\rm NP}) \rangle\right) \left( \langle g_j^{\rm exp}  \rangle - \langle g_j^{\rm th} (\vec{a}_{\rm BGL}, C_{\rm NP}) \rangle\right) \right]_{w-{\rm bin}} \nonumber
\end{align}
where $\hat V$ is the scaled covariance matrix, $N_{w-{\rm bin}}$ are the number of events in the $w$-bin and $\langle g_i^{\rm exp} \rangle$ is the best-fit value of the angular coefficients (see Appendix~\ref{app:statprocedure}  for details) while $\langle g_i^{\rm th} (\vec{a}_{\rm BGL}, C_{\rm NP})\rangle$ are the theoretical expressions, which depend on the form factors ($\vec{a}_{\rm BGL}$) and NP Wilson coefficients ($C_{\rm NP}$). We use the BGL form factors, where the BGL parameters are $\vec{a}_{\rm BGL}=(a^f_n, a^g_n, a^{\mathcal{F}_1}_n, a^{\mathcal{F}_2}_n)$, given in Eq.~\eqref{eq:bglff}. They are described in Appendix~\ref{app:bglformfactors}. Note that as the $\chi^2_{\rm angle}$ term is made from the normalized angular coefficients, the overall normalization constants such as $V_{cb}$ cancel out.
\par
\begin{figure}
\centering
\includegraphics[width=0.9\textwidth]{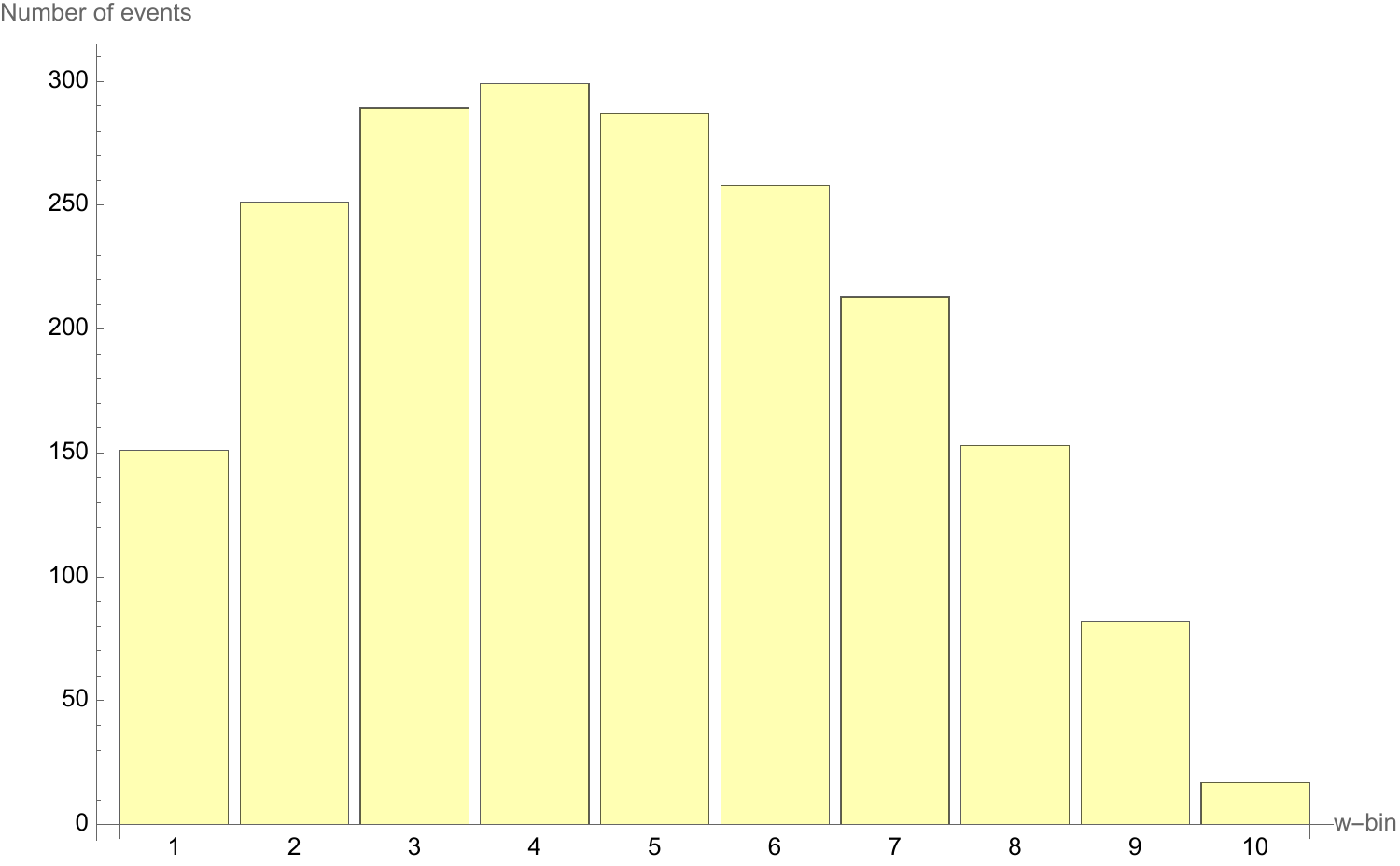}
\caption{\small Histogram of number of events in the 10 $w$-bins, distributed according to the $w$ distribution of the decay rate. The total number of events is 2000.}
\label{fig:histogram}
\end{figure}

As mentioned above, an experimental analysis for this decay has not yet been performed. Belle has around 350 events with 770${\rm fb^{-1}}$ of integrated luminosity, obtained via semileptonic tagging method~\cite{lfubellec}, and about 500 events from the hadronic tagging method \cite{lfubellea}. On the other hand, Belle II also expects to have a larger data sample in a few years (with an improved detector). 
Therefore, we assume that a future combined data sample from Belle and Belle II would have around $2000$ events with hadronic tagging for $\BDstaufull$ decay.
Therefore, we perform the simulation study with 2000 events, whose distribution in the 10 $w$-bins is shown in Figure~\ref{fig:histogram}.
\par
The second term in the $
\chi^2$ comes from the Lattice QCD data. Lattice QCD can compute the values of the form factors in the low $w$ region. 
Fermilab-MILC (FM) and JLQCD collaborations give the values of the four form factors, $F\equiv(f, g, \mathcal{F}_1, \mathcal{F}_2)$, at these three points:  $w_{\rm FM}=(1.03. 1.10, 1.17)$ for Fermilab-MILC collaboration and $w_{\rm JLQCD}=(1.025. 1.060, 1.100)$ for JLQCD~\footnote{{Hereafter, we focus only on the results provided by these two collaborations as the result from~\cite{hpqcd} is given in terms of the HQET form factors and requires an extra transformation.}}. They also provide the $12\times 12$ covariance matrix for them.  Using these lattice data, we fit the BGL parameters, $\vec{a}_{\rm BGL}=(a^f_n, a^g_n, a^{\mathcal{F}_1}_n, a^{\mathcal{F}_2}_n)$, given in Eq.~\eqref{eq:bglff}. The $\chi^2$ can be symbolically written as 
\begin{align}\label{eq:chi2lattice}
\chi^2_{\rm latt}(\vec{a}_{\rm BGL})=(\vec{F}_{\rm BGL}(w_{\rm latt})-\vec{F}_{\rm latt}(w_{\rm latt}))V_{\rm latt}^{-1}(\vec{F}_{\rm BGL}(w_{\rm latt})-\vec{F}_{\rm latt}(w_{\rm latt}))^T,
\end{align}
where ''latt'' is either FM or JLQCD.  The $\vec{F}_{\rm latt}(w_{\rm latt})$ values are the form factors evaluated by the lattice and $V_{\rm latt}$ is the covariance matrix mentioned above. The $\vec{F}_{\rm BGL}(w_{\rm latt})$ values are the form factor expanded in terms of the BGL parameters $\vec{a}_{\rm BGL}$ and evaluated, using  Eq.~\eqref{eq:bglff}, at $w_{\rm latt}$. We keep terms up to $n=2$ in the following. Using the two relations in Eqs.~\eqref{eq:relF1} and \eqref{eq:relF2}, we eliminate $a^{\mathcal{F}_1}_0$ and $a^{\mathcal{F}_2}_2$ and we fit 10 parameters.  
\par 
In addition to constraints from lattice ($\chi^2_{\rm latt}$) and angular distribution data ($\chi^2_{\rm angle}$), we can also use the branching ratio measurement. We use the average of $\BDstaunu$ branching ratio obtained by PDG~\cite{pdg22}, which uses the LHCb~\cite{lhcbbastaunu} and Belle~\cite{bellebdstaunu} measurement of the $\BDstaunu$ branching ratio in its fit. The measurement values are $(1.42 \pm 0.094(\rm stat) \pm 0.140(\rm syst)) \times 10^{-2}$ and $(2.02^{+0.40}_{-0.37}(\rm stat) \pm 0.37 (\rm syst))\times 10^{-2}$, respectively. The PDG average is given by\footnote{The PDG 2024~\cite{pdg24} gives an average value of $(1.45 \pm 0.10)\times 10^{-2}$ for $\BDstaunu$ branching ratio, despite using the same experimental values in its fit. } 
\begin{align}
\mathcal{B}^{\rm exp} (\BDstaunu) = (1.58 \pm 0.09)\times 10^{-2}
\end{align}
leading to a $\chi^2$ contribution
\begin{align}
\chi^2_{\mathcal{B}}(\vec{a}_{\rm BGL},V_{cb}, C_{\rm NP}) = \left(\frac{\mathcal{B}^{\rm th}(\vec{a}_{\rm BGL}, V_{cb}, C_{\rm NP})-0.0158}{0.0009}\right)^2
\end{align}
where $\mathcal{B}^{\rm th}(\vec{a}_{\rm BGL}, V_{cb}, C_{\rm NP})$ is the theoretical prediction of the branching ratio including NP contribution for $\BDstaunu$ case. The SM value of the branching ratio of $\BDstaunu$ obtained using different sets of form factors (Belle, JLQCD and Fermilab-MILC) are:
\begin{align}
\begin{aligned}
\mathcal{B}^{\rm th}(\vec{a}_{\rm Belle}, V_{cb}, C_{\rm NP}=0) &= 0.0142 \\
\mathcal{B}^{\rm th}(\vec{a}_{\rm JLQCD}, V_{cb}, C_{\rm NP}=0) &= 0.0121 \\
\mathcal{B}^{\rm th}(\vec{a}_{\rm Fermilab-MILC}, V_{cb}, C_{\rm NP}=0) &= 0.0116
\end{aligned}
\end{align}
where $V_{cb} = 41.78(70) \times10^{-3}$ is the indirect value obtained by the CKMfitter~\cite{ckmfitter} group, when its directly measured value is not used. The UTfit~\cite{utfit} value is consistent with that of CKMfitter, but we chose the former as it has a larger error, making it a more conservative choice. We see that these values of the branching ratio obtained above fall short of the world average (obtained by experiments) by about $10-20\%$. This might be just another manifestation of the $R(D^{(*)})$ anomaly. 
\par 
To study the effect of $V_{cb}$, we add an additional constraint to our fit using the following $\chi^2$ term
\begin{align}
\chi^2_{V_{cb}}(V_{cb}) = \left( \frac{|V_{cb}|\eta_{\rm EW} - 0.04178*\eta_{\rm EW}}{0.00070} \right)
\end{align}
Finally, the total $\chi^2$ is 
\begin{align}\label{eq:chi2totaltau}
\begin{aligned}
\chi^2_{\rm unbinned}(\vec{a}_{\rm BGL}, V_{cb}, C_{\rm NP})&=\chi^2_{\rm angle}(\vec{a}_{\rm BGL}, C_{\rm NP})+\chi^2_{\rm latt}(\vec{a}_{\rm BGL}) \\
&+\chi^2_{\mathcal{B}}(\vec{a}_{\rm BGL}, V_{cb}, C_{\rm NP}) + \chi^2_{V_{cb}}(V_{cb})
\end{aligned}
\end{align}
\par 
In the following section, we discuss the fit results, with the correlation plots given in Figures~\ref{fig:taucorrelationplots} and~\ref{fig:taucorrelationplots2}. \cn{As mentioned before, the correlation matrix is not sensitive to slight changes in input values of form factors. Therefore, the obtained correlations and sensitivities would not change significantly with actual experimental data.} We only show the fit results for the NP Wilson coefficient $C_{V_R,P,T}$ and $|V_{cb}|$, as the form factor fit values are compatible with those obtained in $\BDslnudpi$ case in~\cite{bdlnumypaper}. This is due to the fact that form factors are mostly constrained by the lattice data, and $\chi^2_{\rm latt}$ is the same in both fits.

\subsection{Fit results}
\cng{Given that we expect little to no correlation between the different Wilson coefficients (as the corresponding operators are sufficiently orthogonal to each other), we assume a single non-zero NP coupling for each fit (i.e. $C_{V_R}\neq 0, C_{P}\neq 0$ or $C_{T}\neq 0$). For each NP coupling, we perform two fits, one assuming that it is real and the other assuming it is imaginary. Let us see the results.}
\subsubsection{Right-handed model: $C_{V_R} \neq 0$}
\textbf{$\mathbf{C_{V_R}}$ is real:}\footnote{The case of imaginary $C_{V_R}$ is not presented, as the corresponding $\chi^2$ is too complex to obtain a minima. However, the expected sensitivity for imaginary $C_{V_R}$ is around the same order as that of imaginary $C_{T}$, as both depend upon almost the same number of observables ($J$-functions), as can be seen from Table~\ref{tab:jfunctioncombinations}.}
\par
The fitted results of real $C_{V_R}$ and $|V_{cb}|$ are given below. We have two sets of fits, one with JLQCD lattice data (Eq~\eqref{eq:rhfitjlqcd}) and one with Fermilab-MILC lattice data (Eq~\eqref{eq:rhfitfermi}). The corresponding correlation plots are shown in Figures~\ref{fig:rhjlqcd} and~\ref{fig:rhfermi}, respectively. In both sets, to study the impact of the $V_{cb}$ constraint, we perform two fits, one by keeping the $\chi^2_{V_{cb}}$ term in Eq.~\eqref{eq:chi2totaltau}, and the other without it. 
\\
Fit results from simultaneous fit of simulated data ($C_{V_R} \neq 0$) and JLQCD lattice data yield:
\begin{align}\label{eq:rhfitjlqcd}
\begin{aligned}
{\rm without} \; \chi^2_{V_{cb}}: 
C_{V_R} &= -0.05 \pm 0.10\\
V_{cb} &= 0.0435 \pm 0.0042 \\
{\rm with} \; \chi^2_{V_{cb}}: 
C_{V_R} &= -0.078 \pm 0.044 \\
V_{cb} &= 0.0421 \pm 0.0007
\end{aligned}
\end{align}
Fit results from simultaneous fit of simulated data ($C_{V_R} \neq 0$) and Fermilab-MILC lattice data yield:
\begin{align}\label{eq:rhfitfermi}
\begin{aligned}
{\rm without} \; \chi^2_{V_{cb}}: 
C_{V_R} &= -0.126 \pm 0.095\\
V_{cb} &= 0.0412 \pm 0.0035 \\
{\rm with} \; \chi^2_{V_{cb}}: 
C_{V_R} &= -0.108 \pm 0.047 \\
V_{cb} &= 0.0420 \pm 0.0007
\end{aligned}
\end{align}
The most interesting result here is the fact that the sensitivity to $C_{V_R}$ improves significantly when we add the constraint from $V_{cb}$ in the $\chi^2$. The statistical uncertainty decreases from $\sim10\%$ to $\sim 4.5\%$, for the same number of events for both lattice datasets. This is also visible in Figures~\ref{fig:rhjlqcd} and~\ref{fig:rhfermi}: the size of dotted curve (without $\chi^2_{V_{cb}}$) is much larger than the solid curve (with $\chi^2_{V_{cb}}$). As noted above, we expect to have about 2000 events for this decay in a few years. Therefore, such an analysis on actual experimental data could yield stringent constraints on right-handed NP in $\BDstaunu$ for the first time. 

\subsubsection{Pseudoscalar model: $C_P \neq 0$}
\textbf{$\mathbf{C_P}$ is real:}
\par
\begin{figure}[H]
\centering
\begin{subfigure}{0.4\textwidth}
\centering
\includegraphics[width=\textwidth]{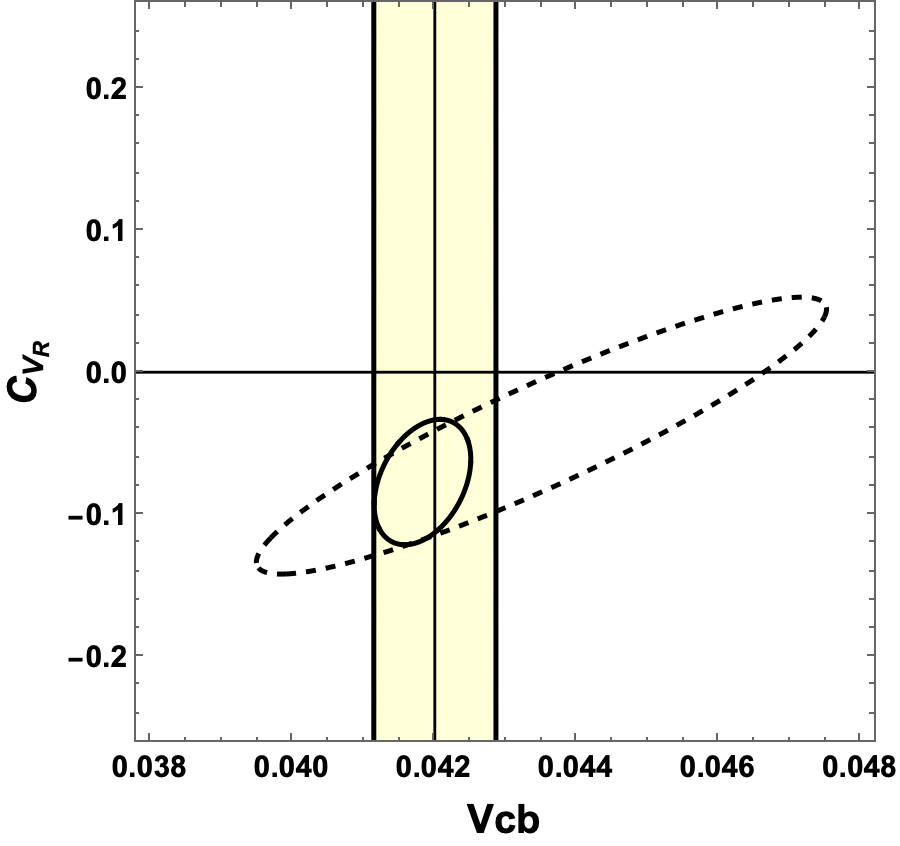}
\caption{\small $C_{V_R} - V_{cb}$ correlation plot with JLQCD lattice data}
\label{fig:rhjlqcd}
\end{subfigure}
\hfill
\centering
\begin{subfigure}{0.4\textwidth}
\centering
\includegraphics[width=\textwidth]{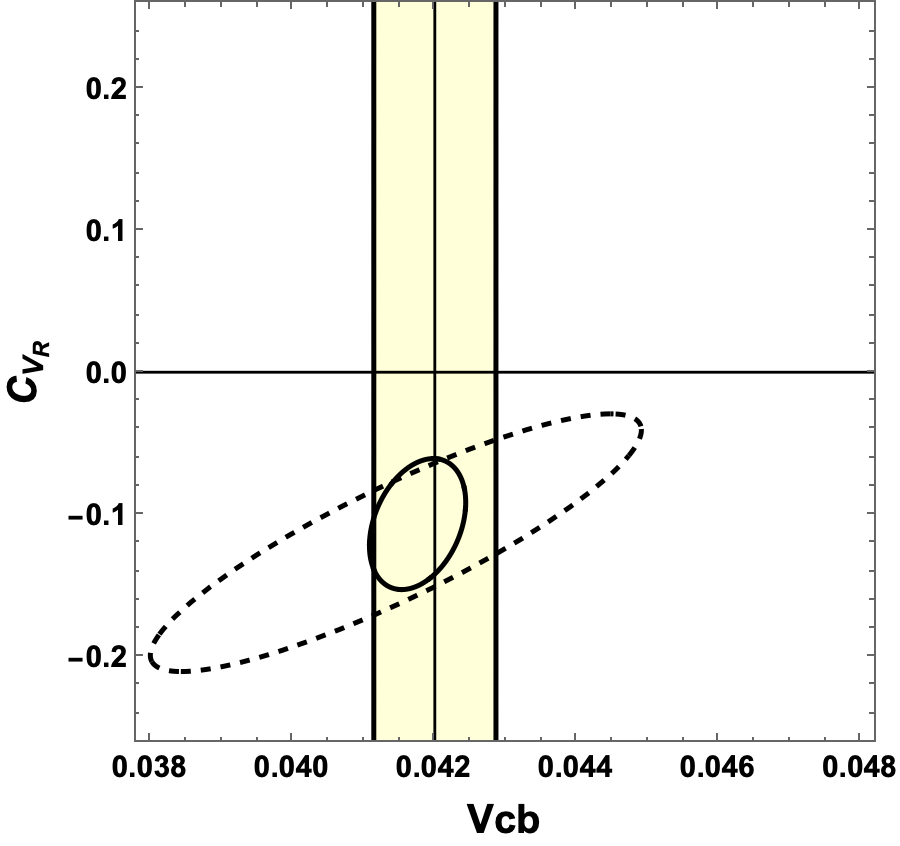}
\caption{\small $C_{V_R} - V_{cb}$ correlation plot with Fermilab-MILC lattice data}
\label{fig:rhfermi}
\end{subfigure}
\hfill
\begin{subfigure}{0.4\textwidth}
\centering
\includegraphics[width=\textwidth]{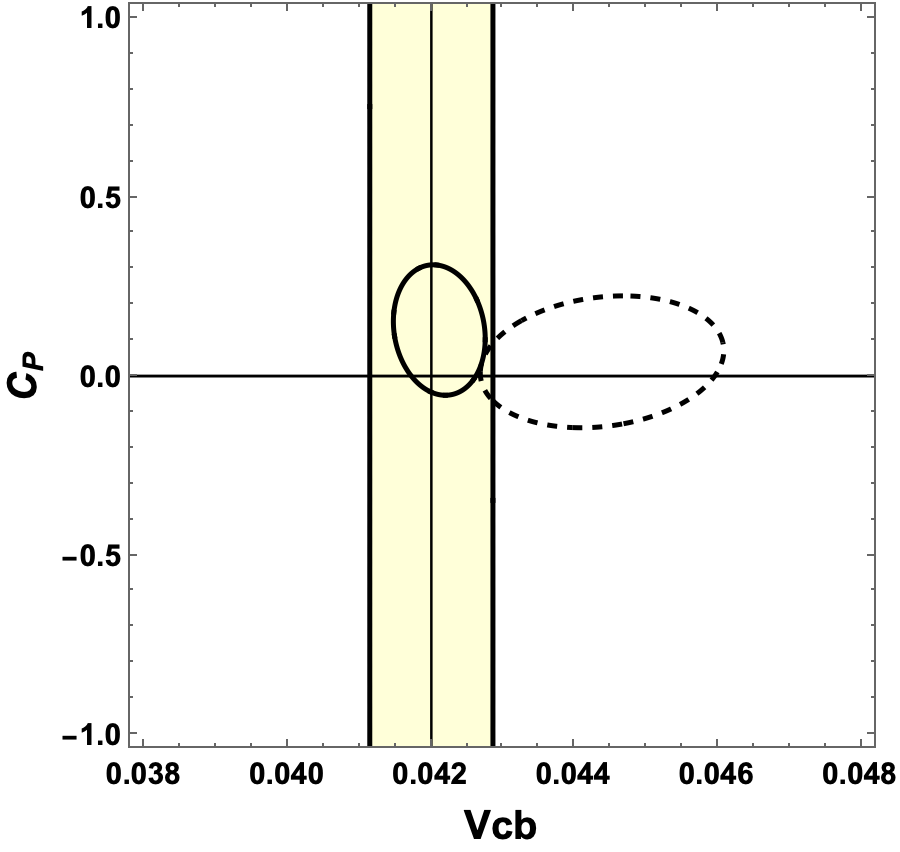}
\caption{\small $C_{P} - V_{cb}$ correlation plot with JLQCD lattice data}
\label{fig:psjlqcd}
\end{subfigure}
\hfill
\begin{subfigure}{0.4\textwidth}
\centering
\includegraphics[width=\textwidth]{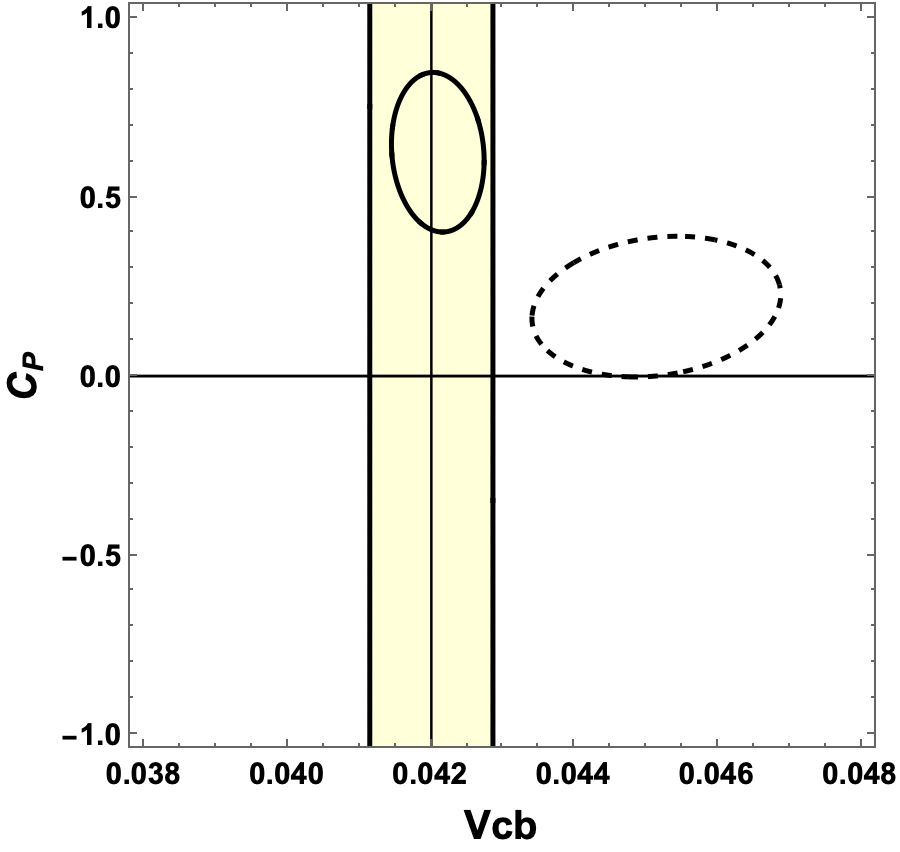}
\caption{\small $C_{P} - V_{cb}$ correlation plot with Fermilab-MILC lattice data}
\label{fig:psfermi}
\end{subfigure}
\caption{\small Correlation plots between $V_{cb}$ and $C_{NP} \; (NP \in \{ RH, P \})$ from the combined fit of simulated data and JLQCD (left column)/Fermilab-MILC (right column) lattice data for the $\BDstaufull$ case with 2000 events. In all the cases, $C_{NP}$ is assumed to be real. The contours indicate a 1$\sigma$ range. The solid contour is obtained by including the $\chi^2_{V_{cb}}$ constraint (as shown in Eq.~\eqref{eq:chi2totaltau}), while the dotted contour is obtained without this term in the global fit. The yellow region within the vertical lines shows the $1\sigma$ interval of $V_{cb}$ obtained by the CKMfitter group~\cite{ckmfitter}.}
\label{fig:taucorrelationplots}
\end{figure}

The fitted results of real $C_{P}$ and $|V_{cb}|$ are given below. As before, we have two sets of fits, one with JLQCD lattice data (Eq~\eqref{eq:psfitjlqcd}) and Fermilab-MILC lattice data (Eq~\eqref{eq:psfitfermi}). The corresponding correlation plots are shown in Figures~\ref{fig:psjlqcd} and~\ref{fig:psfermi}, respectively. In both sets, to study the impact of $V_{cb}$ constraint, we perform two fits, one keeping the $\chi^2_{V_{cb}}$ term in Eq.~\eqref{eq:chi2totaltau}, and the other without it. 
\\
Fit results from a simultaneous fit to simulated data ($C_P \neq 0$) and JLQCD lattice data:
\begin{align}\label{eq:psfitjlqcd}
\begin{aligned}
{\rm without} \; \chi^2_{V_{cb}}: 
C_{P} &= 0.05 \pm 0.19\\
V_{cb} &= 0.0443 \pm 0.0018 \\
{\rm with} \; \chi^2_{V_{cb}}: 
C_{P} &= 0.25 \pm 0.21 \\
V_{cb} &= 0.0423 \pm 0.0006
\end{aligned}
\end{align}
Results from a simultaneous fit to simulated data ($C_P \neq 0$) and Fermilab-MILC lattice data:
\begin{align}\label{eq:psfitfermi}
\begin{aligned}
{\rm without} \; \chi^2_{V_{cb}}: 
C_{P} &= 0.54 \pm 0.28\\
V_{cb} &= 0.0436 \pm 0.0018 \\
{\rm with} \; \chi^2_{V_{cb}}: 
C_{P} &= 0.62 \pm 0.22 \\
V_{cb} &= 0.0423 \pm 0.0007
\end{aligned}
\end{align}
This time, there is not much change in the sensitivity of $C_P$ by adding the $V_{cb}$ constraint; it remains around $20\%$. This follows from the fact that the correlation between these two quantities is much smaller, as can be seen in   Figures~\ref{fig:psjlqcd} and~\ref{fig:psfermi}. 
However, even though we have much fewer events as compared to the $\BDslnu$ case, the sensitivity is comparable, as we now have interference terms of $C_P$ with SM term (i.e. $C_{V_L}$) in the angular observables. This happens thanks to the sizable $\tau$ mass.
\\\\
\textbf{$\mathbf{C_P}$ is imaginary:}
\par 
As can be seen from Table~\ref{tab:jfunctioncombinations}, the imaginary $C_P$ coefficient depends on only one observable ($J_7$). Therefore, the sensitivity to imaginary $C_P$ is quite poor ($O(\sigma_{C_P}) \approx 200\%$), and thus we do not present it here.

\subsubsection{Tensor model: $C_T \neq 0$}
\textbf{$\mathbf{C_T}$ is real:}
\par
\begin{figure}[H]
\centering
\begin{subfigure}{0.4\textwidth}
\centering
\includegraphics[width=\textwidth]{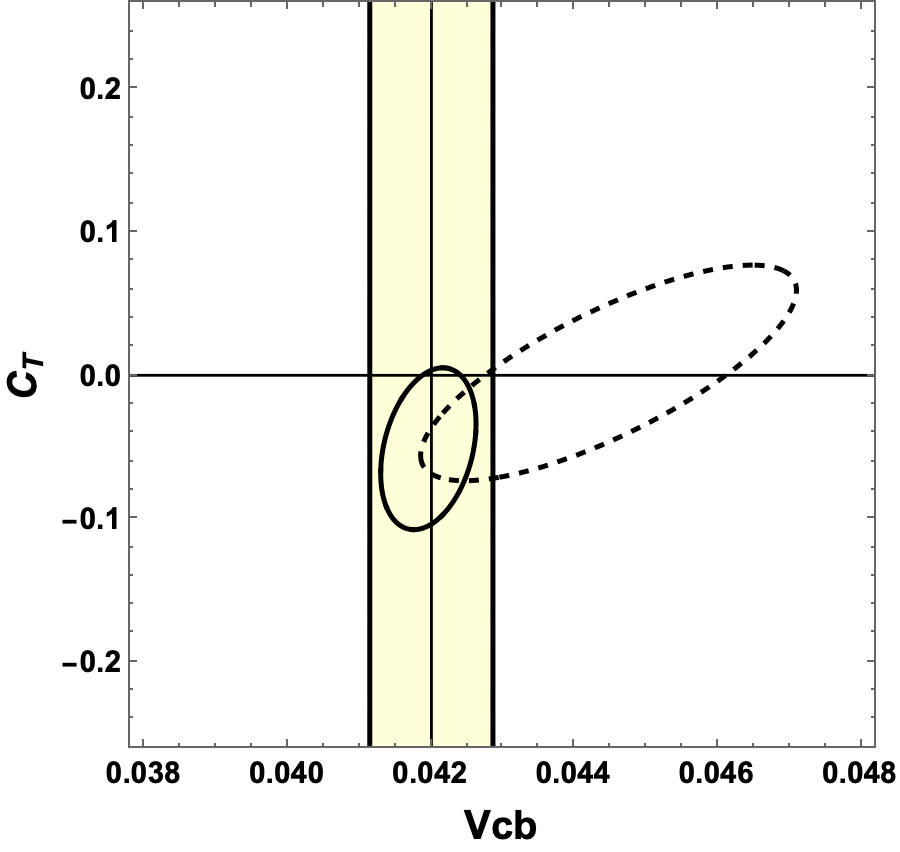}
\caption{\small $C_{T} - V_{cb}$ correlation plot with JLQCD lattice data}
\label{fig:tjlqcd}
\end{subfigure}
\hfill
\begin{subfigure}{0.4\textwidth}
\centering
\includegraphics[width=\textwidth]{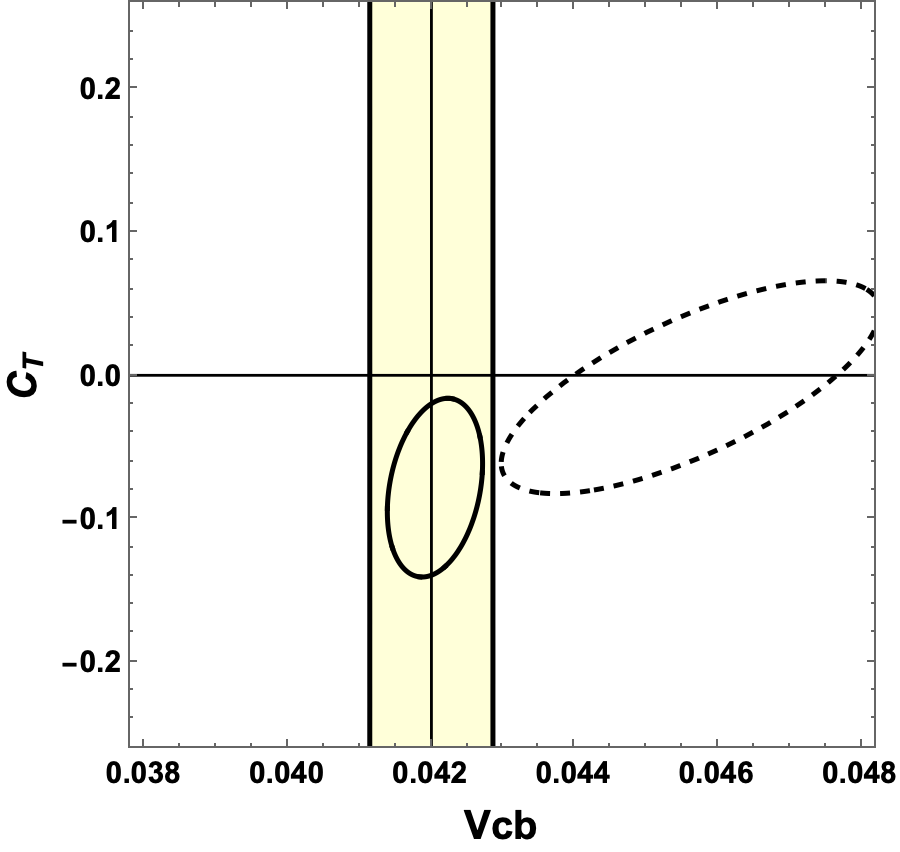}
\caption{\small $C_{T} - V_{cb}$ correlation plot with Fermilab-MILC lattice data}
\label{fig:tfermi}
\end{subfigure}
\hfill
\begin{subfigure}{0.4\textwidth}
\centering
\includegraphics[width=\textwidth]{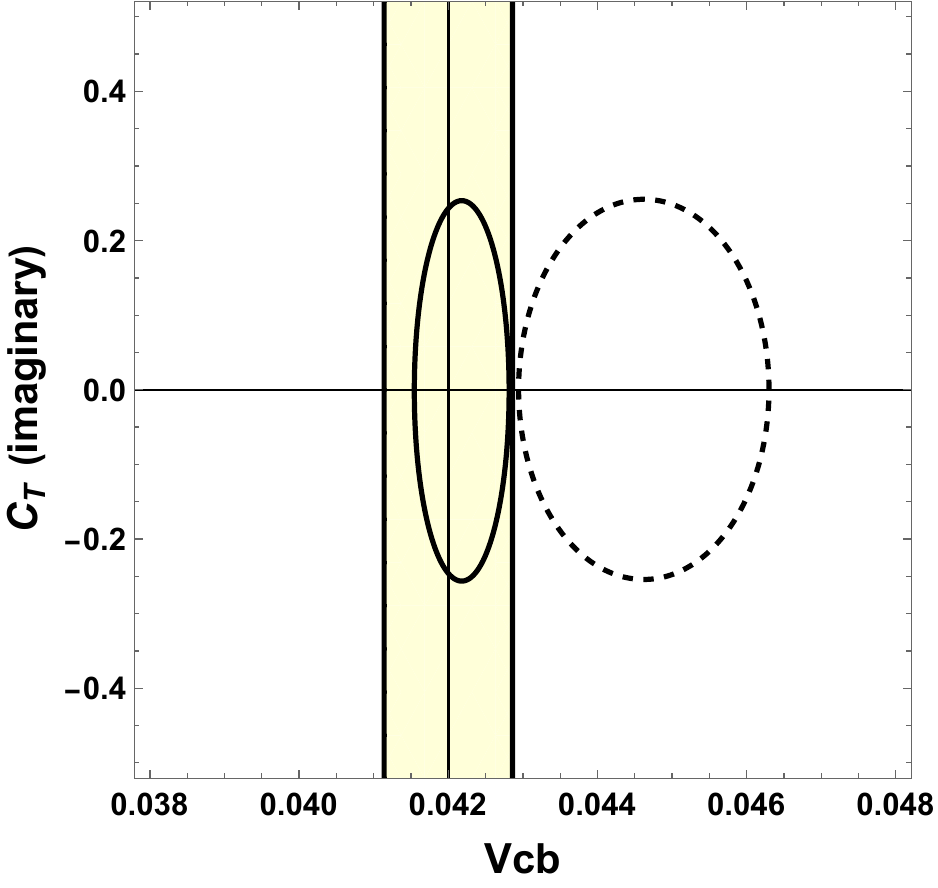}
\caption{\small $C_{V_R} - V_{cb}$ correlation plot with JLQCD lattice data}
\label{fig:timjlqcd}
\end{subfigure}
\hfill
\centering
\begin{subfigure}{0.4\textwidth}
\centering
\includegraphics[width=\textwidth]{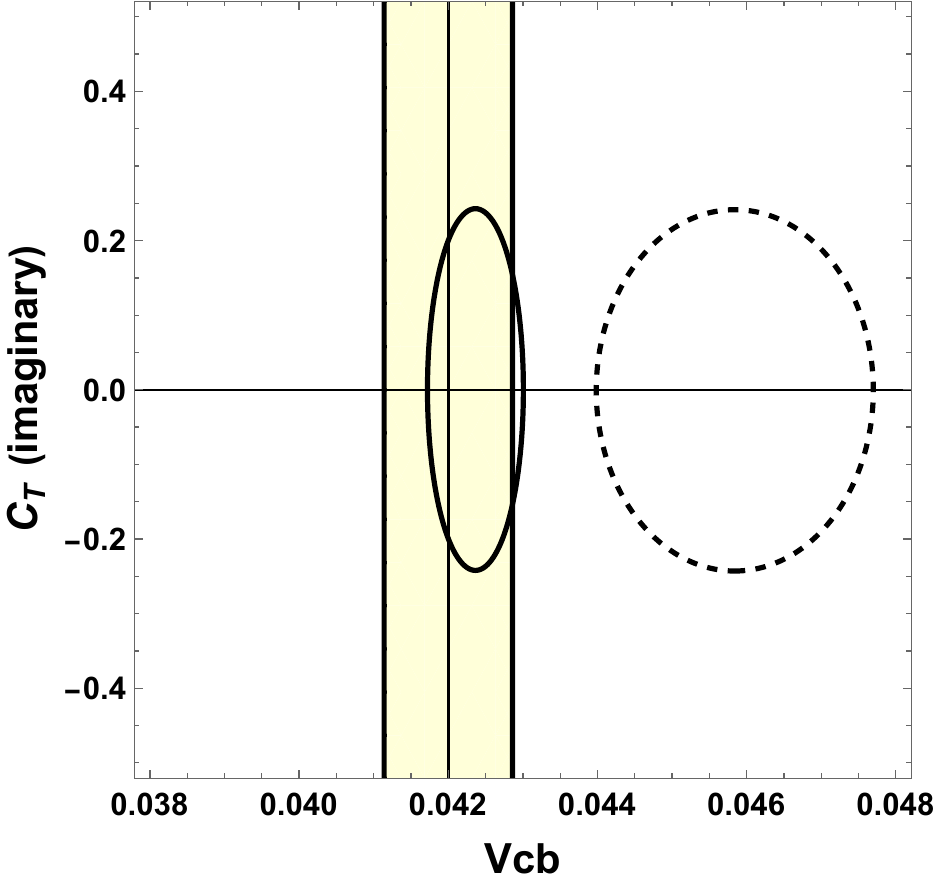}
\caption{\small $C_{V_R} - V_{cb}$ correlation plot with Fermilab-MILC lattice data}
\label{fig:timfermi}
\end{subfigure}
\caption{\small Correlation plots between $V_{cb}$ and $C_{T}$ from the combined fit of simulated data and JLQCD (left column)/Fermilab-MILC (right column) lattice data for $\BDstaufull$  case with 2000 events. In the first row, $C_T$ is assumed to be real, and in the second row, imaginary. The contours indicate a 1$\sigma$ range. The solid contour is obtained by including the $\chi^2_{V_{cb}}$ constraint (as shown in Eq.~\eqref{eq:chi2totaltau}), while the dotted contour is obtained without this term in the global fit. The yellow region within the vertical lines shows the $1\sigma$ interval of $V_{cb}$ obtained by the CKMfitter group~\cite{ckmfitter}.}
\label{fig:taucorrelationplots2}
\end{figure}

The fitted results of real $C_{T}$ and $|V_{cb}|$ are given below. Again, we have two sets of fits, one with JLQCD lattice data (Eq~\eqref{eq:tfitjlqcd}) and one with Fermilab-MILC lattice data (Eq~\eqref{eq:tfitfermi}). The corresponding correlation plots are shown in Figures~\ref{fig:tjlqcd} and~\ref{fig:tfermi}, respectively. In both sets, to study the impact of the $V_{cb}$ constraint, we perform two fits, one keeping the $\chi^2_{V_{cb}}$ term in Eq.~\eqref{eq:chi2totaltau}, and the other without it. 
\\
Results from a simultaneous fit to simulated data ($C_T \neq 0$) and JLQCD lattice data:
\begin{align}\label{eq:tfitjlqcd}
\begin{aligned}
{\rm without} \; \chi^2_{V_{cb}}: 
C_{T} &= 0.012 \pm 0.075\\
V_{cb} &= 0.0449 \pm 0.0027 \\
{\rm with} \; \chi^2_{V_{cb}}: 
C_{T} &= -0.042 \pm 0.055 \\
V_{cb} &= 0.0423 \pm 0.0007
\end{aligned}
\end{align}
Fit results from a simultaneous fit of simulated data ($C_T \neq 0$) and Fermilab-MILC lattice data:
\begin{align}\label{eq:tfitfermi}
\begin{aligned}
{\rm without} \; \chi^2_{V_{cb}}: 
C_{T} &= -0.039 \pm 0.084\\
V_{cb} &= 0.0451 \pm 0.0027 \\
{\rm with} \; \chi^2_{V_{cb}}: 
C_{T} &= -0.058 \pm 0.061 \\
V_{cb} &= 0.0423 \pm 0.0007
\end{aligned}
\end{align}
Here, we notice a modest decrease in statistical uncertainty in $C_T$, from $\sim8 \%$ to $\sim6 \%$, by adding the $V_{cb}$ constraint. In addition, as for the pseudoscalar case, the interference term between $C_T$ and SM ($C_{V_L}$) survives thanks to the large $\tau$ mass. Therefore, even with fewer events, the sensitivity of $C_T$ is better than that of in the $\BDslnu$ case, highlighting the importance of studying the $\BDstaunu$ decay. We also find a stronger correlation between $C_T$ and $V_{cb}$, compared to the $\BDslnu$ case, which plays a role in providing better sensitivity.
\\\\
\noindent
\textbf{$\mathbf{C_T}$ is imaginary:}
\par
The fitted results of imaginary $C_{T}$ and $|V_{cb}|$ are given below. As before, we have two sets of fits, one with JLQCD lattice data (Eq~\eqref{eq:tfitimjlqcd}) and one with Fermilab-MILC lattice data (Eq~\eqref{eq:tfitimfermi}). The corresponding correlation plots are shown in Figures~\ref{fig:timjlqcd} and~\ref{fig:timfermi}, respectively. In both sets, to study the impact of the $V_{cb}$ constraint, we perform two fits, one by keeping the $\chi^2_{V_{cb}}$ term in Eq.~\eqref{eq:chi2totaltau}, and the other without it. 
\\
Fit results from simultaneous fit of simulated data (imaginary $C_T \neq 0$) and JLQCD lattice data:
\begin{align}\label{eq:tfitimjlqcd}
\begin{aligned}
{\rm without} \; \chi^2_{V_{cb}}: 
C_{T} &= (0.00 \pm 0.26)i\\
V_{cb} &= 0.0446 \pm 0.0017 \\
{\rm with} \; \chi^2_{V_{cb}}: 
C_{T} &= (0.00 \pm 0.26)i \\
V_{cb} &= 0.0424 \pm 0.0006
\end{aligned}
\end{align}
Fit results from simultaneous fit of simulated data (imaginary $C_T \neq 0$) and Fermilab-MILC lattice data:
\begin{align}\label{eq:tfitimfermi}
\begin{aligned}
{\rm without} \; \chi^2_{V_{cb}}: 
C_{T} &= (0.00 \pm 0.24)i\\
V_{cb} & = 0.0458 \pm 0.0019 \\
{\rm with} \; \chi^2_{V_{cb}}: 
C_{T} & = (0.00 \pm 0.24)i \\
V_{cb} & = 0.0426 \pm 0.0006
\end{aligned}
\end{align}\cng{Thus, in summary, with 2000 events, the best sensitivities can be obtained for the right-handed and tensor cases. The statistical error on real $C_{V_R}$ can reach $4.5\%$ and is about $5.5\%$ for real $C_T$. With large Belle II datasets, such analyses would soon be possible and would provide one of the most stringent tests of SM. }
\par
Finally, a global comment on $V_{cb}$ (applicable to all of the above fits): in most fits the $|V_{cb}|$ value is higher than the global average of the exclusive measurement when the constraint from $V_{cb}$ is not included. This feature is due to the branching ratio constraint. Recall that in the previous section, we mentioned that our theoretical prediction and the world average value of the branching fraction for $\BDstaunu$ differ by about $10-20\%$ - in fact, the theoretical value is on the smaller side. This forces the fit to increase the $|V_{cb}|$ value to compensate for the branching ratio in order to match the world average value.

\subsection{Effect of neglecting muon mass}
In the calculations above, we neglected the muon, $\mu$, mass, treating $e$ and $\mu$ on the same footing. To study the impact of this approximation, we performed the above calculation including the $\mu$ mass and compared it with the above results. The following were our findings: 
\\
1. There was a negligible change in sensitivity to NP coefficients.
\\
2. The change in $\chi^2 / {\rm dof}$ when we include the muon mass was around $1-3\%$. 
\\
3. In the plot of $\frac{d\Gamma}{dE_\ell}$ as a function of $E_\ell$, the addition of the $\mu$ mass shifts the starting point of the curve slightly to the right (by $M_\mu - M_e$). However, in $B$ factories, the lepton detection turns on above 600 MeV in the lab frame and becomes fully efficient above 800 MeV. Therefore, in such low-energy regions, experiments cannot detect the $\mu$. 

\par
Thus, in our sensitivity study, we have neglected the $\mu$ mass, as this approximation has a negligible effect on our results. Yet, for completeness, the theoretical formulas are given with the lepton mass.

\section{Conclusions}\label{sec:conclusions}
The study of semileptonic $B$ decays is crucial in particle physics for several reasons. Since the final state includes leptons, which interact only through the electroweak force and thus can be factorized from the hadronic part, the theoretical predictions from the SM are less affected by QCD effects, making comparisons between experimental results and theoretical predictions more precise.  In addition, they allow us to determine the CKM matrix parameters important for the study of the SM and to search for physics beyond the SM. Finally, they provide us with a ground to test for lepton flavour universality, which is inherently linked to the non-Abelian $SU(2)_L$ gauge symmetry of the SM.
\par
Motivated by these factors, in this work, we investigated the potential of an angular analysis of $\BDstaunu $ decay to search for NP in light of the new lattice QCD data on the $B \to D^*$ form factors. A complete angular distribution study cannot be performed experimentally, as the $\tau$ is not directly observed in experiment; its presence can only be detected via their decay products.  However, since their leptonic decays contain one or more neutrinos, it is not possible to obtain the $\tau$'s direction. Therefore, we cannot measure the $\tau$'s helicity angles.  Yet, searching for NP is important in this decay, as several models predict the presence of NP couplings in the $\BDstaunu$ decay to explain the $\mathcal{R}(D^{(*)})$ anomaly.  Therefore, we study the angular distribution of the $\BDstaunu$ decay by including the subsequent $\tlnn$ decay, where $\linemu$. Since e and $\mu$ are observed directly in experiments, we can measure its helicity angles and study the angular distribution to search for NP.
\par 
Starting with an effective Hamiltonian containing different NP operators (left/right-handed, pseudoscalar and tensor), we first computed the complete angular distribution of $\BDstaufull$ decay by separating the leptonic and hadronic parts and evaluating them in separate reference frames (following the method in \cite{korner}). As the $\tau$ is observed in experiments only via its decay products, we consider the case with the $\tlnn$ decay, where $\linemu$. To write an observable angular distribution, inspired by the method introduced in ~\cite{bdltaunudatta}, we write the $\ell$ angles and energy in the $W$ rest frame and not in the $\tau$ rest frame. \cng{To obtain the $W$ rest frame, we need the $B$ frame, which is accessible only in $B$-factories, such as SuperKEKB. Thus, this is an interesting physics case to study for Belle II.} 

Navigating through the challenging phase space computation, we also describe the technique to calculate the $\ell$ energy range, which is non-trivial due to the interdependence of variables $q^2$, $E_\ell$ and $\cos\theta_{\tau\ell}$. We find that there are two separate integration ranges for $E_\ell$ and $\cos\theta_{\tau\ell}$, which leads to two sets of $J$-functions (angular coefficients in decay distribution). We confirm these limits with ~\cite{emulimitschina}~\cite{emulimitsspanishpaper}, where a similar calculation was performed.  Finally, we checked the distribution by matching the $q^2$ distribution of $\BDstaunu$ obtained from our result with the one given in the literature. 
\par
In the last part of this article, we investigated the possibility of an unbinned fit for $\BDstaufull$. Here, since there is no experimental data available for this decay, we use the fit values of $B \to D^*$ form factors obtained from $\BDslnudpi$ decay from~\cite{bellefittechnique} and insert them in the angular distribution for $\BDstaufull$  to generate simulated data with 2000 events. We obtain the sensitivities and correlations of NP couplings with other measurements, which can be obtained via this simulated dataset.
\par
We use the simulated data along with the JLQCD/Fermilab lattice data and branching ratio constraints from the $\BDstaunu$ branching ratio global average to perform a sensitivity study for right-handed, pseudoscalar and tensor NP models. In addition, we also add an additional constraint from the global $V_{cb}$ fit value obtained from CKMfitter~\cite{ckmfitter}. For the right-handed case, we find that including the $V_{cb}$ constraint improves the sensitivity to real $C_{V_R}$, and reduces the statistical error from $\sim 10\%$ to $\sim 4.5\%$. For the pseudoscalar case, the $V_{cb}$ constraint has little effect, as the correlation between real $C_P$ and $V_{cb}$ is small. Its statistical error remains around $\sim 20\%$. However, compared to the $\BDslnu$ case~\cite{bdlnumypaper}, we have better sensitivity here (despite having fewer events), as many terms that were suppressed by the lepton mass are no longer suppressed. A similar conclusion holds for the tensor case, with statistical error on real $C_T$ decreasing from $\sim 8\%$ to around $\sim 6\%$ with the $V_{cb}$ constraint. On the other hand, the statistical error on imaginary $C_T$ is around $\sim 20\%$. Finally,
in most of the cases, we find an unusually large value of $|V_{cb}|$ whenever we do not include the $V_{cb}$ constraint. This is attributed to the fact that the theoretically predicted branching ratio of $\BDstaunu$ is smaller than its experimental world average value, so the fit pulls the value of $V_{cb}$ to bring the branching ratio closer to the global value.
\section*{Acknowledgments}
This work was financially supported by the
U.S. National Science Foundation under Grant No.\ PHY-2310627 (BB) and PHY-2309937 (AD). T.E.B thanks the DOE Office of High Energy
Physics for support through DOE grant DE-SC0010504. TK and EK thank F. Le Diberder for his advice and discussions, and their work was in part supported by TYN (the Toshiko-Yuasa Network). TK also express his gratitude to Karim Trabelsi for insightful discussions. 
\section*{Appendix}
\appendix

\section{Canonical form factor parametrization and helicity amplitudes}\label{app:tradformfactors}
The helicity amplitudes are projections of the $\bar{B} \to D^*$ matrix elements on the $W$ boson polarization vectors. To write down the expression of helicity amplitudes, we first express the matrix elements in terms of the form factors.
\par
The vector and axial vector operator matrix elements can be written as
\begin{align}
\langle D^* (p_{D^*},\eds) | \bar{c} \gamma_\mu b | \bar{B}(p_B) \rangle &= -i \epsilon_{\mu\nu\rho\sigma} \eds^{*\nu}p_B^{\rho}p_{D^*}^{\sigma} \frac{2 V(q^2)}{M_B + M_{D^*}}, \label{eq:vectormatrixcanonical} \\
\langle D^* (p_{D^*}\eds) | \bar{c} \gamma_\mu \gamma_5 b | \bar{B}(p_B) \rangle &= \epsilon^{*}_{D^*,\mu} (M_B + M_{D^*}) A_1(q^2) \nonumber \\
&- (p_B+p_{D^*})_{\mu}(\eds^*.q) \frac{A_2(q^2)}{M_B + M_{D^*}} \nonumber \\
&- q_\mu (\eds^*.q) \frac{2 M_{D^*}}{q^2} [A_3(q^2) - A_0(q^2)] ,\label{eq:axialvectormatrixcanonical}
\end{align}
where
\begin{align}
A_3(q^2) = \frac{M_B + M_{D^*}}{2M_{D^*}} A_1(q^2) - \frac{M_B - M_{D^*}}{2M_{D^*}} A_2(q^2),
\end{align}
$q^\mu = p_B^\mu - p_{D^*}^\mu$, $\eds$ is the polarisation vector of $D^*$, and $p_{B(D^*)}$ is the momentum of $B(D^*)$. $V$ is the vector form factor, while $A_{0,1,2,3}$ are the axial-vector form factors (since $A_3(0) = A_0(0)$, only 4 form factors are independent.) 
\par
The tensor matrix element is parameterized as
\begin{align}\label{eq:tensormatrixcanonical}
\begin{aligned}
\langle D^* (p_{D^*},\eds) | \bar{c} \sigma_{\mu\nu}  b | \bar{B}(p_B) \rangle &= \epsilon_{\mu\nu\rho\sigma} \Bigg\{ -\eds^{*\rho} (p_B + p_{D^*})^{\sigma} T_1(q^2) \\
&+ \eds^{*\rho}q^{\sigma} \frac{m_B^2 - m_{D^*}^2}{q^2} [T_1(q^2) - T_2(q^2)] \\
&+ 2 \frac{(\eds^* . q)}{q^2} p_B^{\rho} p_{D^*}^{\sigma} \left[T_1(q^2) - T_2(q^2) - \frac{q^2}{m_B^2 - m_{D^*}^2}T_3(q^2) \right] \Bigg\},
\end{aligned}
\end{align}
where $T_{1,2,3}$ are the tensor form factors. The pseudo tensor matrix elements can be related to the tensor matrix elements by the relation $\bar{c}\sigma_{\mu\nu} \gamma^5 b = -\frac{i}{2} \epsilon_{\mu\nu\alpha\beta} \bar{c}\sigma^{\alpha\beta}b$ with the convention $\epsilon_{0123} = 1$.

\mycomment{
We write the tensor matrix element in another form, which will be useful later:
\begin{align}
\langle D^* (p_{D^*},\eds) | \bar{c} \sigma_{\mu\nu} q^{\nu} b | \bar{B}(p_B) \rangle  &= \epsilon_{\mu\nu\rho\sigma} \eds^{*\nu} p_B^{\rho}p_{D^*}^{\sigma} 2 T_1(q^2) \label{eq:tensorqcanonical}\\
\langle D^* (p_{D^*},\eds) | \bar{c} \sigma_{\mu\nu} \gamma_5 q^{\nu} b | \bar{B}(p_B) \rangle &= -\Bigg[ (m_B^2 - m_{D^*}^2) \epsilon^{*}_{D^*\mu} - (\eds^*.q)(p_B + p_{D^*})_{\mu} \Bigg] T_2(q^2) \nonumber \\
& -(\eds^*.q) \Bigg[ q_{\mu} - \frac{q^2}{m_B^2 - m_{D^*}^2} (p_B + p_{D^*})_{\mu} \Bigg] T_3(q^2). \label{eq:pseudotensorqcanonical}
\end{align}
\par 
}

We do not need to define a separate form factor for the pseudoscalar form factor, as it can be expressed in terms of $A_0(q^2)$ (as derived in Appendix~\ref{app:tradformfactors}):
\begin{align}\label{eq:pseudoscalarmatrixcanonical}
\langle D^* (p_{D^*},\eds) | \bar{c} \gamma_5 b | \bar{B} (p_B) \rangle = -(\eds^*.q) \frac{2 M_{D^*}}{m_b + m_c} A_0(q^2),
\end{align}
where $m_{b(c)}$ is the $b(c)$-quark mass. 
\par 
Having defined the matrix elements,  using Eq.~\eqref{eq:hadhelamp}, we define the hadronic helicity amplitudes as
\begin{align}\label{eq:helamp}
\begin{aligned}
H_{\rm V}^\pm(q^2)  &\equiv H_{\rm V_L,\pm}^{\pm}(q^2) = -H_{\rm V_R,\mp}^{\mp}(q^2) \\
&= (M_B + M_{D^*}) A_1(q^2) \mp \frac{\sqrt{\lambda_{D^*}(q^2)}}{M_B + M_{D^*}}V(q^2), \\
H_{\rm V}^0(q^2) &\equiv H_{\rm V_L,0}^{0}(q^2) = - H_{\rm V_R,0}^{0}(q^2) \\
&= \frac{M_B + M_{D^*}}{2 M_{D^*}\sqrt{q^2}} \left[ -(M_B^2 - M_{D^*}^2 - q^2) A_1(q^2) + \frac{\lambda_{D^*}(q^2)}{(M_B + M_{D^*})^2}A_2(q^2) \right], \\
H_{P}(q^2) &\equiv H_{P}^0(q^2) \\
&= -\frac{\sqrt{\lambda_{D^*}(q^2)}}{M_B + m_c} A_0(q^2), \\
H_{T}^\pm(q^2) &\equiv \pm H_{\rm{T},\pm t}^{\pm}(q^2) \\
&= \frac{1}{\sqrt{q^2}} \bigg[ \sqrt{\lambda_{D^*}(q^2)} T_1(q^2) \pm (M_B^2 - M_{D^*}^2) T_2(q^2) \bigg], \\
H_{T}^0(q^2) &\equiv H^0_{\rm{T},+-}(q^2) = H^0_{\rm{T},0t}(q^2) \\
&= \frac{1}{2 M_{D^*}} \bigg[ (M_B^2 + 3 M_{D^*}^2 - q^2) T_2(q^2) - \frac{\lambda_{D^*}(q^2) }{(M_B^2 - M_{D^*}^2)}T_3(q^2) \bigg]. \\
\end{aligned}
\end{align}
These expressions agree with the results in \cite{london} and \cite{helampref2} while we find an overall sign difference in $H_{T,0}$ with respect to~\cite{tensorff}.

\section{BGL form factor parametrisation}\label{app:bglformfactors}

In the lattice results, the form factors are often parameterized using the BGL parametrization \cite{bgl}.  BGL parametrization uses a series expansion in terms of a small parameter $z$ derived from the conformal mapping of the kinematical variable $w$. This mapping allows $z$ to map the kinematic range of $q^2$ (or $w$) to a small interval,  improving the convergence properties of the series.  In addition, with more parameters,  BGL parametrization can accommodate a wider range of possible behaviours of the form factors, leading to potentially better fits to experimental data. Thus, in this work, we work with the BGL parametrization. The canonical form factors are related to them as follows: 
\begin{align}\label{eq:bgltostdff}
\begin{aligned}
g&=\frac{2}{M_B+M_{D^*}}V, \\
f&=(M_B+M_{D^*})A_1, \\
\mathcal{F}_1 &= \frac{1}{2M_{D^*}}\left[ (M_B^2-M_{D^*}^2-q^2)(M_B+M_{D^*}) A_1  -  
\frac{4M_B^2|{ \vec{p}}_{D^*}|^2}{ M_B+M_{D^*}} A_2 \right],  \\
\mathcal{F}_2 &= 2A_0 ,
\end{aligned}
\end{align}
where the $q^2$ (or $w$) dependence of the form factors is implicit. 
By using these definitions, the helicity amplitudes given in Eq.~\eqref{eq:helamp} can now be written in a very simple manner: 
\begin{align} \label{eq:helampbgl}
\begin{aligned}
H_{V}^\pm(w) &= f\mp g M_B |{\vec{p}_{D^*}}| , \\
H_{V}^0(w) &=\frac{\mathcal{F}_1 }{M_{B}\sqrt{1-2wr+r^2}}, \\
H_{P}(w)&=  -\frac{\mathcal{F}_2 M_B |{ \vec{p}_{D^*}}|}{m_b + m_c}, \\
H_{T}^\pm(w)&=\frac{\pm f (m_b - m_c) + g M_B |{ \vec{p}_{D^*}}|(m_b + m_c)}{ M_{B}\sqrt{1-2wr+r^2} }, \\
H_{T}^0(w) &= \frac{-(m_b-m_c)(-\mathcal{F}_1 (M_B^2 - M_{D^*}^2) + 2 \mathcal{F}_2 M_B^2 |{ \vec{p}_{D^*}}|^2 ) }{(M_B^2 - M_{D^*}^2) {(M_{B}^2 + M_{D^*}^2 - 2M_{B}M_{D^*} w  )}},
\end{aligned}
\end{align}
where $m_{b(c)}$ is the $b(c)$-quark mass. Note that the pseudoscalar and tensor form factors are reduced to the above four form factors thanks to the relations given in Appendix~\ref{app:tradformfactors}.
\par
The momentum dependence of these form factors is given in a $z-$ expansion, where
\begin{equation}
z \equiv \frac{\sqrt{w+1}-\sqrt{2}}{\sqrt{w+1}+\sqrt{2}},
\end{equation}
and the form factors are
\begin{align} \label{eq:bglff}
\begin{aligned}
g(z) &= \frac{1}{P_{1^-}(z)\phi_g(z)} \sum_{n=0}^{\infty} a^g_n z_n,  \quad \quad \;
f(z) = \frac{1}{P_{1^+}(z)\phi_f(z)} \sum_{n=0}^{\infty} a^f_n z^n, \\
\mathcal{F}_1(z) &= \frac{1}{P_{1^+}(z)\phi_{\mathcal{F}_1}(z)} \sum_{n=0}^{\infty} a^{\mathcal{F}_1}_n z_n,  \quad
\mathcal{F}_2(z) = \frac{1}{P_{0^-}(z)\phi_{\mathcal{F}_2}(z)} \sum_{n=0}^{\infty} a^{\mathcal{F}_2}_n z_n .
\end{aligned}
\end{align}
Note that the expansion coefficients must satisfy the unitarity conditions: 
\begin{align}\label{unitarity}
\sum_{n=0}^\infty  (a^g_n)^2<1, \quad 
\sum_{n=0}^\infty  (a^f_n)^2+(a^{\mathcal{F}_1}_n)^2<1, \quad 
\sum_{n=0}^\infty  (a^{\mathcal{F}_2}_n)^2<1. 
\end{align}
From Eq.~\eqref{eq:bgltostdff}, we find that the form factors are not completely independent.

\begin{table}[t]
\begin{center}
\renewcommand{\arraystretch}{1.4}
\begin{tabular}{|c||c|c|}
\hline 
Type & $B_c^{(*)}$ mass (GeV) & $\chi^{T, L}_{1^{\pm}}$ \\ \hline\hline
$f$, $\mathcal{F}_1$ & 6.739, 6.750, 7.145, 7.150  & $3.894\times 10^{-4}\  {\rm GeV}^{-2}$\\ \hline
$g$ & 6.329, 6.920, 7.020, 7.280 &$5.131\times 10^{-4}\  {\rm GeV}^{-2} $ \\ \hline
$\mathcal{F}_2$ & 6.275, 6.842, 7.250 &$1.9421\times 10^{-2}$  \\\hline
\end{tabular}
\end{center}
\caption{\small The input parameters for $P_{1^-, 1^+, 0^-}$ and $\phi_{g, f, \mathcal{F}_1, \mathcal{F}_2}$ functions used in the BGL form factors given in Eq.~\eqref{eq:bglff}. We take the values used  in~\cite{fermilab}}
\label{table:bglinput}
\end{table}

At the zero-recoil limit, i.e. $|{ \vec{p}_{D^*}}|=0$ $(w_{\rm min}=1)$, we find that both $f$ and $\mathcal{F}_1$ are written only by the $A_1$ form factor and are related as  
\begin{align}\label{eq:relF1}
\mathcal{F}_1(0)=(M_B-M_{D^*})f(0) .
\end{align}
On the other hand, at the maximum recoil, $q^2=0$ $(w_{\rm max}=\frac{M_B^2+M_{D^*}^2}{2M_BM_{D^*}})$, we have a condition, $A_0=\frac{M_B+M_{D^*}}{2M_{D^*}}A_1-\frac{M_B-M_{D^*}}{2M_{D^*}}A_2$, to avoid the $q^2$ pole in the definition of the axial-vector operator matrix element \cite{tensorff}. This leads to 

\begin{align}\label{eq:relF2}
\mathcal{F}_1(z(w_{\rm max}))=\frac{(M_B^2-M_{D^*}^2)}{2}\mathcal{F}_2(z(w_{\rm max})) .
\end{align}

Taking into account these relations, we can eliminate two expansion parameters. We choose to eliminate the lowest order $\mathcal{F}_1$ constant, i.e $a^{\mathcal{F}_1}_0$ and the highest order $\mathcal{F}_2$ constant $a^{\mathcal{F}_2}_{j_{\rm max}}$.

The functions $P_{1^-, 1^+, 0^-}$ and $\phi_{g, f, \mathcal{F}_1, \mathcal{F}_2}$, whose expressions can be found in the original article~\cite{bgl}, contain several input parameters. We follow the reference~\cite{fermilab} (originally obtained in~\cite{gambino}) and use the values given in Table~\ref{table:bglinput} as input parameters\footnote{These values are very different from the ones used in { \cite{belle19} } which actually picks the values from \cite{bgl}. The choice of these values affects the relationship between the form factors and its expansion coefficients (see Eq.~\eqref{eq:bglff}). }, with $n_I=2.6$, $M_B = 5.280$ GeV and $M_{D^*} = 2.010$ GeV.

\section{Statistical procedure}\label{app:statprocedure}
\subsection{Maximum likelihood estimation}
Starting with the full dataset obtained from experiment, one can employ the maximum likelihood estimation to obtain the best-fit values of parameters. However,  dealing with full statistical data is a tedious task. Therefore, for our purposes, we will describe a method which allows us to obtain the covariance matrix corresponding to a given number of events.
\par 
Let us define the normalized PDF of our model as 
\begin{align}
\hat{f}_{\vec{v}}(\vec{x})
\end{align}
where the hat means that the PDF is normalized.  In the decays we studied, this PDF corresponds to the normalized angular distributions (for example, the one given in Eq.~\eqref{eq:normpdftau}.) We have
\begin{itemize}
\item $\vec{v}$: These are the physical parameters in the PDF, for example, the form factors and the NP Wilson coefficients. We estimate these parameters using maximum likelihood estimation.
\item$\vec{x}$ refers to the phase space parameters.
\end{itemize}
We know that we can obtain the best-fit value of the parameters $\vec{v}$ by the maximum likelihood estimator $\vec{v}^*$ obtained by maximizing the likelihood $L$.  It is defined as the product of the PDF of each event.  However, it is often more convenient to obtain $\vec{v}^*$ by minimizing the log-likelihood $\mathcal{L} = \ln L$ (as taking a log changes the product of PDFs to sum). In particular, when the events are Gaussian distributed, the $\chi^2$ is related to the log-likelihood as
\begin{align} \label{eq:likelihoodchi2}
\chi^2 = -2  \mathcal{L}(\vec{v}), \quad  {\rm where} \quad \mathcal{L} (\vec{v}) = \sum_{i=1}^N \ln\hat{f}_{\vec{v}}(\vec{x}_i) 
\end{align}
where $N$ are the number of events. The maximum likelihood estimator $\vec{v}^*$ is obtained by
\begin{align}
\frac{\partial \mathcal{L} (\vec{v}) }{ \partial \vec{v} } \Bigg|_{\vec{v} = \vec{v}^*} = 0
\end{align}
Finally, it can also be shown that the covariance matrix is given by
\begin{align}\label{eq:covariancematrixdef}
V^{-1}_{ab} \equiv -\frac{\partial^2 \mathcal{L}(\vec{v})}{\partial v_a \partial v_b} \Bigg|_{\vec{v}=\vec{v}^*}
\end{align}
\subsection{Toy Monte Carlo study}
In this section, we introduce the \emph{Toy Monte Carlo method}, which can be used for a simple sensitivity study. It provides the achievable statistical precision for a given number of events, provided we have the \emph{truth-level data points} for the parameters of PDF.  These truth-level data points are the best-fit values of the parameters found by experiment.  Using these truth-level data points, we will reconstruct the likelihood and, thus, the covariance matrix.
\par 
The main idea is that since we input these truth-level values, the solution to the maximal likelihood equations should lead to exactly these values if we have large enough (infinite) statistics. The expected statistical error should also approach zero at this limit. However, since the error scales as $1/\sqrt{N_{\rm event}}$, we can rescale the covariance matrix, giving us the statistical error with a limited number of events. 
\par 
First, let us introduce an integral relation: for infinite statistics, the definition of expectation of a function changes as follows - 
the sum of any event can be replaced by an integral over phase space
\begin{align}
\frac{1}{N_{\rm event}} \sum_{i=1}^{N_{\rm event}} [\;] \to \int \hat{f}_{\vec{v}=\vec{v}^*}(\vec{x}) [\;] d\vec{x}
\end{align}
where $\hat{f}_{\vec{v}=\vec{v}^*}$ is the PDF obtained by the truth-level data points. The bracket can contain any function whose expectation value is of interest.  Inserting $\ln\hat{f}_{\vec{v}}$ in this bracket, we obtain the log-likelihood (as defined in Eq.~\eqref{eq:likelihoodchi2}) written in integral form as follows:
\begin{align}\label{eq:likelihoodintegral}
\mathcal{L}(\vec{v}) = N_{\rm event}  \int \hat{f}_{\vec{v}=\vec{v}^*}(\vec{x}) \ln\hat{f}_{\vec{v}} (\vec{x}) d\vec{x}
\end{align}
Now, the covariance matrix is obtained by inserting the likelihood from Eq~\eqref{eq:likelihoodintegral} in Eq.~\eqref{eq:covariancematrixdef}:
\begin{align}\label{eq:covmatrixgeneral}
V^{-1}_{ij} = - N_{\rm event}\int \frac{\partial^2 \hat{f}_{\vec{v}} (\vec{x})}{\partial v_i \partial v_j} \Bigg|_{\vec{v} = \vec{v}^*} d\vec{x} + N_{\rm event}\int \frac{\partial \hat{f}_{\vec{v}}(\vec{x})}{\partial v_i} \Bigg|_{\vec{v} = \vec{v}^*}  \frac{\partial \hat{f}_{\vec{v}}(\vec{x})} {\partial v_i} \Bigg|_{\vec{v} = \vec{v}^*} \frac{1}{\hat{f}_{\vec{v} = \vec{v}^*}} d\vec{x}
\end{align}
where we recall that the truth-level data points $\vec{v}^*$ are the best-fit values obtained experimentally, and $\hat{f}_{\vec{v}}(\vec{x})$ is the normalised angular distribution and $\vec{x}$ denotes the phase space variables. By adjusting the number of events $N_{\rm event}$, we can obtain the expected statistical error of the parameters $\vec{v}$ corresponding to those many numbers of events
\par
We note that for $\BDslnu$ decay, the normalized PDF is only linearly dependent on the parameters $\vec{v}$. Thus, removing the double derivative term from Eq.~\ref{eq:covmatrixgeneral}, we find 
\begin{align}
V_{ij}^{-1} = N_{\rm event} \int\  \left.\left(\frac{\partial \hat{f}_{\vec{v}} (\vec{x}) }{\partial v_i}  \frac{\partial \hat{f}_{\vec{v}} (\vec{x})}{\partial v_j}  \frac{1}{\hat{f}_{\vec{v}} (\vec{x})}\right)\right|_{\vec{v}=\vec{v}^*}\   d\vec{x} 
\end{align}

For $\BDslnu$ decay:
\begin{itemize}
\item $\hat{f}$ is the normalized probability distribution function given by Eq.~\eqref{eq:normpdftau}.$dx$ represents integration over the complete phase space, i.e. the three angles that describe the decay. 
\item $\vec{v}=\langle \vec{g}_i \rangle$ is the vector of observables given in Eq.~\eqref{eq:gifunctau}. 
\item $\vec{v}^*=\langle \vec{g}_i \rangle ^{\rm exp}$ is the vector of truth values, which we obtain by inserting the experimentally measured form factors from \cite{belle19} in $\vec{v}$.
\end{itemize}
Finally, using this \emph{simulated data}, we perform a $\chi^2$ fit using $\vec{v}_i$ with our model assumptions, which we call $\vec{v}_{i}^{\rm \; model}$: 
\begin{align}
\chi^2 = \sum_{i,j} (\vec{v}_i^{\rm{ \; model}} - \vec{v}_{i}^{\;*})V^{-1}_{ij}(\vec{v}_j^{\rm{\; model}} - \vec{v}_{j}^{\;*}).
\end{align}
\par
For convenience, we can define a rescaled covariance matrix $\hat{V}$ as 
\begin{align}
\hat{V} =N_{\rm event} V 
\end{align}
It should be noted that since we are not using the actual experimental data, the central values obtained by this method are not exact. However, it gives reliable sensitivity estimates and provides a guide for future theoretical and experimental studies.

\bibliographystyle{JHEP}
\bibliography{bibliographybdstaunu.bib}
\end{document}